\documentclass[12pt]{article}

\usepackage[T1]{fontenc}
\usepackage[utf8]{inputenc}
\usepackage[english]{babel}

\usepackage{amssymb}
\usepackage{amsmath}
\usepackage{amsthm}
\usepackage{cancel}
\usepackage{mathdots}

\usepackage{appendix}

\usepackage{color}

\usepackage{longtable}

\usepackage{mathtools}

\usepackage{enumerate}

\allowdisplaybreaks
\def\proof{\begin{center} {\bf Proof:} \end{center}\vspace{0.5pt}}
\def\QEDclosed{\mbox{\rule[0pt]{1.3ex}{1.3ex}}} 
\def\QED{\QEDclosed} 
\def\endproof{\hspace*{\fill}~\QED\par\endtrivlist\unskip}

\newtheorem{thm}{Theorem}[]
\newtheorem{coro}[thm]{Corollary}

\newtheorem{lemma}[thm]{Lemma}
\newtheorem{conj}[]{Conjecture}

\newtheorem*{problem}{Problem}

\theoremstyle{definition}
\newtheorem*{rmk}{Remark}
\newtheorem{example}{Example}

\usepackage{palatino}
\begin{document}
\title{On deformations of one-dimensional Poisson structures of hydrodynamic type with degenerate metric}
\author{Andrea Savoldi}
    \date{}
    \maketitle
    \vspace{-7mm}
\begin{center}
Department of Mathematical Sciences, Loughborough University \\
Leicestershire LE11 3TU, Loughborough, United Kingdom \\
e-mail: \\[1ex]
\texttt{A.Savoldi@lboro.ac.uk}\\
\end{center}

\bigskip

\begin{abstract}
We provide a complete list of two- and three-component Poisson structures of hydrodynamic type with degenerate metric, and study their homogeneous deformations. In the non-degenerate case any such deformation is trivial, that is, can be obtained via Miura transformation. We demonstrate that in the degenerate case this class of deformations is non-trivial, and depends on a certain number of arbitrary functions. This shows that the second Poisson-Lichnerowicz cohomology group does not vanish.

\bigskip


\bigskip

\end{abstract}

\newpage

\tableofcontents

\section{Introduction}
One-dimensional Poisson structures (also called Hamiltonian operators) of hydrodynamic type were introduced by Dubrovin and Novikov in \cite{DN}. They are defined by
\begin{equation}\label{Ham1}
P^{ij}=g^{ij}({\bf u}) \frac{d}{dx} + b^{ij}_k ({\bf u}) u^k_x,
\end{equation}
where ${\bf u}=(u^1,\ldots, u^n)$ are local coordinates depending on the spatial variable $x$, $i,j,k=1,\ldots,n$, and $u^k_x$ means the derivative of $u^k$ with respect to $x$.
The requirement that \eqref{Ham1} defines a Poisson bracket, by
$$
\{ F,G \} =\int \frac{\delta F}{\delta u^i} P^{ij} \frac{\delta G}{\delta u^j},
$$
imposes certain constraints on the coefficients $g^{ij}$ and $b^{ij}_k$.
In the non-degenerate case, that is when $\det g^{ij}\neq0$, Dubrovin and Novikov proved that \eqref{Ham1} determines a Poisson structure if and only if $g^{ij}$ is a pseudo-Riemannian flat metric and the symbols $\Gamma^{i}_{jk}=-g_{jm}b^{mi}_{k}$ define a connection compatible with the metric $g_{ij}$ (where $g^{i m}g_{mj}=\delta^{i}_{j}$). As direct consequence of this result, any non-degenerate Hamiltonian operator of the form \eqref{Ham1} can be reduced to constant coefficients by local change of coordinates. The theory of non-degenerate Hamiltonian structures of such type was thoroughly investigated in the last three decades, and Hamiltonian operators of the form \eqref{Ham1} have subsequently been generalised in a whole variety of different ways (degenerate \cite{Grinberg,B1,B2}, non-homogeneous \cite{MF1,Mok4}, higher-order \cite{DN1,P,P1,Do,FPV}, multi-dimensional \cite{DN1,DN2,Mok1,Mok4,Mok2,FLS} and non-local \cite{MF,F,Mok3,MF1}, see \cite{Mok} for further review).

In the framework of the theory of Frobenius manifolds \cite{Dubrovin1,DZ1,DZ}, Dubrovin conjectured the triviality of homogeneous formal deformations of structures \eqref{Ham1}. The problem formulated by Dubrovin can be stated as follows. Let us consider a Poisson manifold $M$ endowed whit a Poisson structure of hydrodynamic type (bivector) $P_0$ which satisfies the Jacobi condition written in terms of the Schouten bracket on the algebra of multivector fields on $M$, that is $[P_0,P_0]=0.$
A deformation of order $k$ of a Poisson bivector $P_0$ is a formal series
$$
P_{\epsilon}=P_0 + \epsilon P_1 + \epsilon^2 P_2 + \ldots
$$
in the space of bivector fields on $M$ satisfying the condition $[P_{\epsilon}, P_{\epsilon}]=\mathcal{O}(\epsilon^k)$ for any value of the parameter $\epsilon$. In particular, if $[P_{\epsilon}, P_{\epsilon}]=0$, we say that $P_{\epsilon}$ is a deformation of $P_0$. A deformation (of order $k$) is trivial if there exists a Miura transformation $\phi_\epsilon: M \rightarrow M$,
$$
\phi_{\epsilon}= \sum_{m=0}^{\infty} \epsilon^{m}\phi_m,
$$
which pulls back $P_{\epsilon}$ to $P_0$, that is $P_{\epsilon}=\phi_{\epsilon *} P_0$.
Introducing the following gradation in the space of differential polynomials,
$$
\mathrm{deg} (f({\bf u}))=0, \quad \mathrm{deg} \left(\frac{d^k {\bf u}}{dx^k} \right)=k,
$$
the allowed deformations $P_{\epsilon}$ are formal series of the form
$$
P_k =\sum_{s=0}^{k+1} A_s({\bf u}, {\bf u}_x, \ldots, {\bf u}_{k+1}) \frac{d^{k+1-s}}{dx^{k+1-s}},
$$
where the entries of the $ n \times n$ matrices $A_s$ are homogenous polynomials of degree $s$ in the $x$-derivative, namely $\mathrm{deg}(A_s)=s$.
\begin{problem}
Does there exists a Miura transformation that brings the homogeneous deformation $P_{\epsilon}$ to $P_0$?
\end{problem}

This problem can be reformulated in cohomological terms. Indeed, triviality of deformations is equivalent to the vanishing of the second cohomology group in the Poisson-Lichnerowicz cohomology \cite{L}.
Getzler \cite{Ge} and independently Degiovanni, Sciacca and Magri \cite{De} solved Dubro\-vin's conjecture proving that this cohomology group is trivial (in particular, Getzler proved that all the positive integer cohomology groups are trivial).

In subsequent years, the theory of deformations for Poisson structures of hydrodynamic type has been developed especially in the framework of bi-Hamiltonian structures by several authors (see for instance \cite{DZ, Lor, LZ1, LZ2, DLZ, Ba, AL, Bu, CPS}) and recently the first result has been published in the two-dimensional case \cite{C}.

All the above results have been obtained assuming that the metric $g$ which defines the Poisson structure \eqref{Ham1} is \emph{non-degenerate}. To the best of our knowledge the deformation theory for Poisson structures of hydrodynamic type with \emph{degenerate} metric has not been developed yet.
Our main aim is to investigate what happens in the degenerate framework.

\subsection{Summary of the main results}
In this paper, we give a complete list of two- and three-component Poisson structures of hydrodynamic type with degenerate metric (Theorems \ref{thm2cmpt} and \ref{thm3cmpts}). 
For instance, in two-component case, any of these structures can be brought to one of the following canonical forms
$$
P^{(1)}_0=
\begin{pmatrix}
d_x& 0 \\
0 & 0
\end{pmatrix},
\quad
P^{(2)}_0=
\begin{pmatrix}
d_x & -\frac{u^2_x}{u^1} \\
\frac{u^2_x}{u^1} & 0
\end{pmatrix},
$$
where $d_x=\frac{d}{dx}$. We show that in two-component case first- and second-order deformations of such structures are not trivial, that is, they cannot be eliminated by Miura transformations, and we prove that they depend on a certain number of arbitrary functions of the variable $u^2$.

\begin{thm}\label{coho_2cmpts}
Up Miura-type transformations, the following hold:
\begin{itemize}
\item first-order deformations  of  $P^{(1)}_0$ depend on 2 functions of $u^2$, and
second-order deformations on 6 functions  of $u^2$; 
\item first-order deformations of  $P^{(2)}_0$ depend on 1 function of $u^2$, and
second-order deformations on 2 functions of $u^2$.
\end{itemize}
\end{thm}

In three-component case, we provide some examples of non-trivial first-order deformations (as we will see, in this case second-order deformations involve too many unknown functions, and the computations become very hard), focusing on the Poisson structures given by
$$
P_0^{(3)}=
\begin{pmatrix}
0 & u^3_x & 0\\
-u^3_x & 0 & 0\\
0 & 0 & 0
\end{pmatrix},
\quad
P_0^{(4)}=
\begin{pmatrix}
d_x & 0 & 0\\
0 & 0 & 0\\
0 & 0 & 0
\end{pmatrix},
\quad
P_0^{(5)}=
\begin{pmatrix}
d_x & 0 & 0\\
0 & d_x & 0\\
0 & 0 & 0
\end{pmatrix}.
$$

In particular, our results imply that the first homogeneous component of the second Poisson-Lichnerowicz cohomology group for the structures $P^{(i)}_0$, for $i=1,\ldots5$, does not vanish. This implies that the second cohomology group for such degenerate structures is not trivial, contrary to what happens in the non-degenerate case  \cite{Ge,De} .

This paper is organised as follows.
The first section is devoted to the theory of Poisson structures with degenerate metric, where we recall the main results due to Grinberg \cite{Grinberg} and Bogoyavlensky \cite{B1,B2}, and in \ref{sect_class} we give the complete classification of two- and three-component degenerate Hamiltonian operators (three-component case is fully analysed in Appendix A). Deformations of degenerate Poisson structures are described in Section \ref{sect_deformations}. The results we list in this section are proved in Appendix C, leading to the proof of Theorem \ref{coho_2cmpts}. Finally, Appendix B is a quick recall of the tools we used in computations.

\section{Poisson structures of hydrodynamic type with degenerate metric}\label{sect_degenerate}
In this work, we are interested in the degenerate case, that is, when the determinant of the differential-geometric object $g^{ij}$ which describes the Hamiltonian operator \eqref{Ham1} is zero. Some results about this class of structures were announced for the first time by Grinberg in 1985, in a short communication \cite{Grinberg}, and later investigated by  Bogoyavlenskij \cite{B1,B2}.

The first important result of Grinberg is stated in the following theorem.
\begin{thm}[\cite{Grinberg}]
Operator \eqref{Ham1} defines a Hamiltonian structure if and only if the pair $(g,b)$ satisfies the following conditions
\begin{subequations}\label{1D_cond}
\begin{equation}\label{G1}
g^{ij}=g^{ji},
\end{equation}
\begin{equation}\label{G2}
\frac{\partial g^{ij}}{\partial u^k}=b^{ij}_k+b^{ji}_k,
\end{equation}
\begin{equation}\label{G3}
g^{tk} b^{ji}_t=g^{tj} b^{ki}_t,
\end{equation}
\begin{equation}\label{G4}
b^{ij}_t b^{tk}_r-b^{ik}_t b^{tj}_r=g^{ti} \left( \frac{\partial b^{jk}_r}{\partial u^t} - \frac{\partial b^{jk}_t}{\partial u^r}\right),
\end{equation}
\begin{equation}\label{G5}
\sum_{(i,j,k)} \left[ \left( \frac{\partial b^{ij}_t}{\partial u^q} - \frac{\partial b^{ij}_q}{\partial u^t}\right) b^{tk}_r+\left( \frac{\partial b^{ij}_t}{\partial u^r} - \frac{\partial b^{ij}_r}{\partial u^t}\right) b^{tk}_q \right]=0,
\end{equation}
\end{subequations}
where $\sum_{(i,j,k)}$ means cyclic summation over $i,j,k$.
\end{thm}
The proof of this statement is a direct computation of the skew-symmetry and the Jacobi identity.

In the degenerate situation, up to now a fully geometric interpretation of these equations is not clear: what is known is that the kernel of $g$ defines an integrable distribution \cite{Grinberg}. Bogoyavlensky found some tensor invariant objects for degenerate Hamiltonian structures \cite{B1,B2}, and proved that any solution of equations \eqref{G1}--\eqref{G5} in the case of ${\rm rank} (g^{ij})=m<n$ locally has the form
$$
g^{ij} ({\bf u})=\sum_{\alpha,\beta=1}^m c^{\alpha \beta} U^i_{\alpha} ({\bf u}) U^j_{\beta} ({\bf u}),
\quad
b^{ij}_k ({\bf u})=\sum_{\alpha,\beta=1}^m c^{\alpha \beta} U^i_{\alpha} ({\bf u}) \frac{\partial U^j_{\beta} ({\bf u})}{\partial u^k} + T^{ij}_k ({\bf u}),
$$
where $c^{\alpha \beta} $ are constant coefficients, $U_1({\bf u}), \ldots, U_m({\bf u})$ are commuting vector fields on $M^n$, and the symbols $T^{ij}_k({\bf u})$ form a certain $(2,1)$-tensor which satisfies
$$
T^{ij}_k=-T^{ji}_k, \quad T^{ik}_m g^{mj}=0,
$$
plus extra conditions.

Unfortunately, even if this result simplifies the analysis of Grinberg's conditions \eqref{1D_cond}, to the best of our knowledge there is no classification of such structures in the literature.
Our first aim is to obtain this classification up to three-component case.

In the non-degenerate situation, there always exists a system of coordinates where the pair $(g,b)$ assumes constant form. In general, this is not true, but a weaker result holds:
\begin{thm}[\cite{Grinberg}]
Suppose that \eqref{Ham1} defines a $n$-component Hamiltonian operator, and $\mathrm{rank} (g^{ij})=m \le n$. Then $g^{ij}$ can be reduced to a constant form.
\end{thm}
Thought we can easily classify all possible canonical forms for degenerate constant metrics, the symbols $b$ are no longer defined through $g$. Fixing $g$, the coefficients $b$ can be found solving equations \eqref{1D_cond}.

In her paper, Grinberg gives a description of two- and three-component degenerate Hamiltonian operators (one-component is trivial), without explicitly writing out the canonical forms that such operators take. Here, starting from her results, we list all the possible canonical forms they may assume, up to arbitrary changes of dependent variables.

\subsection{Admissible changes of coordinates}\label{sect_change}

We recall that under arbitrary changes of $u^i$, the components of $g^{ij}$ transform as a $(2,0)$-tensor, i.e. a contravariant metric, while the objects $b^{ij}_k$ transform according to the following rule
\begin{equation}\label{b_transf}
b^{lr}_s(\tilde{{\bf u}})=\frac{\partial \tilde{u}^l}{\partial u^i} \frac{\partial \tilde{u}^r}{\partial u^j} \frac{\partial u^k}{\partial \tilde{u}^s} b^{ij}_k({\bf u})
+\frac{\partial \tilde{u}^l}{\partial u^i} \frac{\partial^2 \tilde{u}^r}{\partial u^j \partial u^k} \frac{\partial u^k}{\partial \tilde{u}^s} g^{ij}({\bf u}).
\end{equation}
In matrix notation, let us consider the operator \eqref{Ham1} as $P=G d_x +B$, where $G^{ij}=g^{ij}$ and $B^{ij}=b^{ij}_k u^k_x$. Since $P$ is a $(2,0)$-tensor, it transforms as $J P J^t$, thus we have
$$
\tilde{P}=J P J^t= J (G d_x) J^t + J B J^t=J G J^t d_x+ J G (J^t)_x+ J B J^t.
$$
Thus, the non-tensorial part of the transformation rule \eqref{b_transf} corresponds to
\begin{equation}\label{b_transf_matrix}
J G (J^t)_x.
\end{equation}

Once the metric is fixed and Grinberg's conditions are solved, in order to reduce them to canonical forms we need a change of coordinates which preserves the form of the metric. Following \cite{Grinberg}, this class of transformations is called \emph{admissible}. Unfortunately, under admissible change of coordinates, the symbols $b^{ij}_k$ do not transform like components of a $(2,1)$-tensor (as we can see in the example below), so we have to be careful with the transformation rules.

For simplicity of the computations, at the beginning we classify Poisson structures under changes of coordinates that both preserve the form of the metric, and transform $b^{ij}_k$ as a tensor. Secondly, we will check whether the structures we have obtained are equivalent under admissible transformations.

\begin{example}\label{example1}
Let us consider the degenerate metric
\begin{equation}\label{metricEx}
g^{ij}=G=
\begin{pmatrix}
0 & 1 & 0\\
1 & 0 & 0\\
0 & 0 & 0
\end{pmatrix}.
\end{equation}
One can easily see that the transformation given by
\begin{equation}\label{transf1}
v^1=u^3 u^1, \quad v^2=\frac{u^2}{u^3}, \quad v^3=u^3
\end{equation}
is an admissible transformation. Since the symbols $b^{ij}_k$ transform according to \eqref{b_transf}, the non-tensorial part of such transformation is given by
\begin{equation}\label{tensorial2}
\sum_{i,j,k}\frac{\partial v^l}{\partial u^i} \frac{\partial^2 v^r}{\partial u^j \partial u^k} \frac{\partial u^k}{\partial v^s} g^{ij}({\bf u}).
\end{equation}
If the symbols $b^{ij}_k$ transform like components of a $(2,1)$-tensor,
then \eqref{tensorial2} has to vanish. Let us compute the inverse transformation:
$$
u^1=\frac{v^1}{v^3}, \quad u^2=v^3 v^2, \quad u^3=v^3.
$$
For instance, let us look at \eqref{tensorial2} in the case where $l=2$, $r=1$ and $s=3$. Considering the fact that the only non-vanishing elements of the metric are $g^{12}=g^{21}=1$, \eqref{tensorial2} reads
$$
\sum_k \left(\frac{\partial v^2}{\partial u^2} \frac{\partial^2 v^1}{\partial u^1 \partial u^k} \frac{\partial u^k}{\partial v^3} +\frac{\partial v^2}{\partial u^1} \frac{\partial^2 v^1}{\partial u^2 \partial u^k} \frac{\partial u^k}{\partial v^3}\right)= \frac{\partial v^2}{\partial u^2} \frac{\partial^2 v^1}{\partial u^1 \partial u^3} \frac{\partial u^3}{\partial v^3}=\frac{1}{u^3} \neq0.
$$
\end{example}

In order to consider changes of coordinates which both preserve the metric $g^{ij}$ and transform the object $b^{ij}_k$ as a tensor, we need to restrict to a subclass of admissible transformations: we have to require that \eqref{b_transf_matrix} vanishes.

\begin{lemma}\label{lemma1}
Suppose that $0<{\rm rank} (g^{ij})=m<n$. Among all the possible transformations which preserve the form of the constant metric $g^{ij}$, those which transform the symbols $b^{ij}_m$ as component of $(2,1)$-tensor must be of the form
\begin{equation}\label{change_Adm}
\tilde{u}^i= c^i_1 u^1+\ldots c^i_m u^m +F^i(u^{m+1},\ldots, u^n), \quad i=1,\ldots, n,
\end{equation}
where $c^{i}_j$ are constants and $c^{i}_j=0$ for $i \in \{m+1,\ldots, n\}$.
\end{lemma}
Of course, the requirement of admissibility imposes some further restrictions on the coefficients $c^{i}_j$.
\begin{rmk}
In the case where ${\rm rank} (g^{ij})=0$, the objects $b^{ij}_k$ always transform as a tensor. Indeed, as said above, under arbitrary changes of the variables $u^1,\ldots,u^n$, the coefficients $b^{ij}_k$ transform according to \eqref{b_transf}. If $g^{ij}$ identically vanishes, then this is exactly the transformation rule for a $(2,1)$-tensor.
\end{rmk}

%%%%
%%%%

\begin{center} {\bf Proof of Lemma \ref{lemma1}:} \end{center}\vspace{0.5pt}

Suppose ${\rm rank} (g^{ij}) = m$, $0<m<n$. Working in the coordinate system where the metric assumes constant coefficients form, without any loss of generality we can consider $g^{ij}$ of the form
$$
g^{ij}=G=
\left(
\begin{array}{c|c}
A& 0\\
\hline
0 & 0
\end{array}
\right),
$$
where $A$ is a non-degenerate ($\det A\neq0$)  $m\times m$ symmetric matrix  with constant coefficients.

If we consider a generic change of coordinates of the form $\tilde{u}^i=\tilde{u}^i(u^1,\ldots,u^n)$, the metric $G$ transforms as
$
\tilde{G}=J G J^{t},
$
where $J^{ij}=\frac{\partial \tilde{u}^i}{\partial u^j}$ is the Jacobian of the change of coordinates, $\det J \neq0$.
Let us write $J$ in block form, namely
\begin{equation}\label{JacMatrix}
J=
\left(
\begin{array}{c|c}
J^{11} & J^{12}\\
\hline
J^{21} & J^{22} 
\end{array}
\right)
=
\left(
\begin{array}{ccc|cccc}
\frac{\partial \tilde{u}^1}{\partial u^1} & \cdots & \frac{\partial \tilde{u}^1}{\partial u^m} & \frac{\partial \tilde{u}^1}{\partial u^{m+1}} & \cdots & \frac{\partial \tilde{u}^1}{\partial u^n} \\
\vdots & \ddots& \vdots & \vdots & \ddots& \vdots \\
\frac{\partial \tilde{u}^m}{\partial u^1} & \cdots & \frac{\partial \tilde{u}^m}{\partial u^m} & \frac{\partial \tilde{u}^m}{\partial u^{m+1}} & \cdots & \frac{\partial \tilde{u}^1}{\partial u^n} \\
\hline
\frac{\partial \tilde{u}^{m+1}}{\partial u^1} & \cdots & \frac{\partial \tilde{u}^{m+1}}{\partial u^m} & \frac{\partial \tilde{u}^{m+1}}{\partial u^{m+1}} & \cdots & \frac{\partial \tilde{u}^{m+1}}{\partial u^n} \\
\vdots & \ddots& \vdots & \vdots & \ddots& \vdots \\
\frac{\partial \tilde{u}^n}{\partial u^1} & \cdots & \frac{\partial \tilde{u}^n}{\partial u^m} & \frac{\partial \tilde{u}^n}{\partial u^{m+1}} & \cdots & \frac{\partial \tilde{u}^n}{\partial u^n} 
\end{array}\right).
\end{equation}
Thus, we have
$$
\left(
\begin{array}{c|c}
\tilde{A}& 0\\
\hline
0 & 0
\end{array}
\right)
=
\left(
\begin{array}{c|c}
J^{11} & J^{12}\\
\hline
J^{21} & J^{22} 
\end{array}
\right)
\left(
\begin{array}{c|c}
A& 0\\
\hline
0 & 0
\end{array}
\right)
\left(
\begin{array}{c|c}
(J^{11})^t & (J^{21})^t\\
\hline
(J^{12})^t & (J^{22})^t 
\end{array}
\right),
$$
which leads to
$$
\left(
\begin{array}{c|c}
\tilde{A}& 0\\
\hline
0 & 0
\end{array}
\right)
=
\left(
\begin{array}{c|c}
J^{11} A  (J^{11})^t & J^{11} A (J^{21})^t\\
\hline
J^{21} A  (J^{11})^t & J^{21} A (J^{21})^t
\end{array}
\right).
$$
Since we are requiring that this transformation preserves the metric $G$, one can easily see that the condition
\begin{equation}\label{condJac1}
\tilde{A}=J^{11} A  (J^{11})^t
\end{equation}
necessarily implies $\det J^{11} \neq 0$, otherwise we would have $\det \tilde{A}=0$. Thus, $J^{11}$ is invertible, and then condition $J^{11} A (J^{21})^t=0$ reads
$
(J^{21})^t=0,
$
which means that for $j \in \{1,\ldots,m\}$ and $i \in \{m+1,\ldots,n\}$, we have
$$
\frac{\partial \tilde u^i}{\partial u^j}=0,
$$
that is,  $\tilde u^i=\tilde u^i(u^{m+1},\ldots, u^{n})$ for every $i \in \{m+1,\ldots,n\}$. This proves \eqref{change_Adm} in the case where $i\in \{m+1,\ldots,n\}$.

Furthermore, since a transformation which preserves the form of a constant non-degenerate metric must be affine, condition \eqref{condJac1} tells us that the transformation of coordinates $\tilde{u}^1,\ldots, \tilde{u}^m$ has to be affine with respect to the coordinates $u^1,\ldots,u^m$, 
\begin{equation}\label{affine}
\frac{\partial^2 \tilde u^i}{\partial u^j \partial u^k}=0, \quad \forall \, i,j,k \in \{1,\ldots,m\},
\end{equation}
or, equivalently,
\begin{equation}\label{affine1}
\tilde u^i= \sum_{k=1}^{m} R^{i}_k(u^{m+1},\ldots,u^n) u^k + F^i(u^m,\ldots,u^n), \quad  i\in \{1,\ldots,m\},
\end{equation}
where $R^i_k$ and $F^i$ are arbitrary functions of $(u^{m+1},\ldots,u^n)$. Let us remark that $R^i_k=(J^{11})^{ik}$, so we will call $J^{11}=R$.
We want to show that the requirement that the symbols $b^{ij}_k$ transform as coefficient of a $(2,1)$-tensor implies that all $R^i_k$ must be constant. 

As said above, in order to obtain the subclass of changes of coordinates which transforms the coefficients $b^{ij}_k$ as  a tensor, we have to require that condition \eqref{b_transf_matrix} holds. Thus, we have
$$
J G  (J^t)_x=
\left(
\begin{array}{c|c}
R & J^{12}\\
\hline
0 & J^{22} 
\end{array}
\right)
\left(
\begin{array}{c|c}
A& 0\\
\hline
0 & 0
\end{array}
\right)
\left(
\begin{array}{c|c}
(R^t)_x & 0\\
\hline
((J^{12})^t)_x & ((J^{22})^t)_x
\end{array}
\right)
\\
=
\left(
\begin{array}{c|c}
R A  (R^t)_x & 0\\
\hline
0 & 0
\end{array}
\right).
$$
Since $R$ and $A$ are invertible matrices, we get $(R^t)_x=0$, which means that the elements of the matrix $R$ must be constant.

\endproof

Let us point out that in general a transformation of the form \eqref{change_Adm} is not admissible: as said above, we need to impose other restrictions on the coefficients $c^{i}_j$. However, this subset of transformations is not empty, since, for instance, setting $c^i_j=\delta^i_j$, \eqref{change_Adm} is effectively an admissible transformation.

If we go back on the Example \ref{example1}, one can easily see that the transformation given by  \eqref{transf1} is not in our class. Indeed, by straightforward computation, one can prove that a generic transformation which preserves the form of metric \eqref{metricEx}, has the form
$$
u^1=F_1(\tilde u^3) \tilde u^1+ F_2(\tilde u^3), \quad u^2= \frac{\tilde u^2}{F_1(\tilde u^3)} + F_3(\tilde u^3), \quad u^3=F_4(\tilde u^3),
$$
or
$$
u^1=F_1(\tilde u^3) \tilde u^2+ F_2(\tilde u^3), \quad u^2= \frac{\tilde u^1}{F_1(\tilde u^3)} + F_3(\tilde u^3), \quad u^3=F_4(\tilde u^3).
$$
Thus, in order to have a transformation in our class, we need to require that $F_1(\tilde u^3)=const$. 

We call \emph{restricted admissible transformation} the subclass of admissible transformations which satisfies relation \eqref{change_Adm}.

\begin{rmk}
Sometimes the restricted class coincides with the more general class of admissible transformations. In this case, in what follows we refer to this class as \emph{(restricted) admissible transformations}.
\end{rmk}

%%%%
%%%%
%%%%

\subsection{Classification results}\label{sect_class}
The complete classification for two- and three-component operators is given in this section. We deal with \emph{non-trivial} structures, that is, we assume \eqref{Ham1} to be not identically zero. We will see that already in the three-component case we have 11 canonical forms (8 if we allow complex changes of coordinates). This suggests that the classification in the four-component case would be not so easy.

Our aim is to solve Grinberg's conditions \eqref{1D_cond}. Let us point out that since we will always work in a coordinate system where the metric can be reduced to constant form, \eqref{G1} is automatically satisfied, while \eqref{G2} implies $b^{ij}_k=-b^{ji}_k$. Thus, $b^{ii}_k=0$ and, without any loss of generality, we can consider as unknowns the coefficients $b^{ij}_k$ where $i<j$.

\subsubsection{Two-component case}
Fixing the number of components, degenerate metrics can be characterised by their rank. In particular, for $n=2$, we have to investigate metrics with ${\rm rank}( g^{ij})=0,1$.
In two-component case, we have only two canonical forms.
\begin{thm}\label{thm2cmpt}
Any non-trivial degenerate two-component Hamiltonian operator of Dubrovin-Novikov type in 1D can be brought, by a change of the dependent variables, to one of the following two canonical forms:
\begin{enumerate}
\item Constant form
\begin{equation}\label{1D_2cmpt_const}
P=\begin{pmatrix}
d_x & 0\\
0 & 0
\end{pmatrix},
\end{equation}
\item Non-constant form
\begin{equation}\label{1D_2cmpt_non_const}
P=
\begin{pmatrix}
d_x & -\dfrac{u^2_x}{u^1}\\
\dfrac{u^2_x}{u^1} & 0
\end{pmatrix}.
\end{equation}
\end{enumerate}
\end{thm}
\proof
If $\mathrm{rank} (g^{ij})=0$ then the Hamiltonian operator is identically zero \cite{Grinberg}. Suppose $\mathrm{rank} (g^{ij})=1$, thus the metric can be reduced by local changes to constant form. Without any loss of generality we can assume 
\begin{equation}\label{metric1}
g^{ij}=
\begin{pmatrix}
1 & 0\\
0 & 0
\end{pmatrix}.
\end{equation}
By straightforward computation we obtain that all $b^{ij}_k$ vanish except $b^{12}_2=-b^{21}_2$, which has to satisfy the condition
$$
\partial_1 b^{12}_2= (b^{12}_2)^2.
$$
If $ b^{12}_2=0$, all the coefficients $b^{ij}_k$ vanish and we have the constant solution \eqref{1D_2cmpt_const}.
Otherwise, for $b^{12}_2\neq0$ we get
$$
b^{12}_2=\frac{1}{f(u^2)-u^1}
$$
for an arbitrary $f(u^2)$.
Applying the (restricted) admissible transformation
$$
\tilde{u}^1 =u^1-f(u^2), \quad \tilde{u}^2 = u^2,
$$
we can reduce $b^{12}_2$ to $-\frac{1}{\tilde{u}^1}$ obtaining \eqref{1D_2cmpt_non_const}.
\endproof

In this case, since a generic admissible transformation for the metric \eqref{metric1} is given by
$
\tilde{u}^1 =u^1+F(u^2),  \tilde{u}^2 = G(u^2)
$,
the classes of restricted and admissible transformations coincide. This implies that the symbols $b^{ij}_k$ transform as tensors under admissible changes of coordinates, and clearly the structures \eqref{1D_2cmpt_const} and \eqref{1D_2cmpt_non_const} cannot be equivalent (since in the first case the coefficients $b^{ij}_k$ vanish, while in  \eqref{1D_2cmpt_non_const} they are non-zero).

%%%%%%
%%%%%%

\subsubsection{Three-component case}
In the three-component case there are three distinct possibilities: $\mathrm{rank}(g^{ij})=0,1,2$.

\begin{thm}\label{thm3cmpts}
Any non-trivial degenerate three-component Hamiltonian operator of Dubrovin-Novikov type in 1D can be brought, by a change of the dependent variables, to one of the following canonical forms:
\begin{itemize}
\item rank(g) = 0:
\end{itemize}
\begin{equation}\label{rank0}
P=\begin{pmatrix}
0 & u^3_x & 0\\
-u^3_x & 0 & 0\\
0 & 0 & 0
\end{pmatrix},
\end{equation}
\begin{itemize}
\item rank(g) = 1:
\end{itemize}
\begin{equation}\label{rank1}
\begin{array}{c}
P=\begin{pmatrix}
d_x & 0 & 0\\
0 & 0 & 0\\
0 & 0 & 0
\end{pmatrix},
\quad
P=\begin{pmatrix}
d_x & u^3_x& 0\\
-u^3_x& 0 & 0\\
0& 0 & 0
\end{pmatrix},
\quad
P=
\begin{pmatrix}
d_x & 0& -\frac{u^3_x}{u^1}\\
0& 0 & 0\\
\frac{u^3_x}{u^1}& 0 & 0
\end{pmatrix},
\\[20pt]
P=
\begin{pmatrix}
d_x & -\frac{u^2_x}{u^1} & -\frac{u^3_x}{u^1}\\
\frac{u^2_x}{u^1} & 0 & 0\\
\frac{u^3_x}{u^1} & 0 & 0
\end{pmatrix},
\end{array}
\end{equation}
\begin{itemize}
\item rank(g)=2:
\end{itemize}
\begin{equation}\label{rank2}
\begin{array}{c}
P=\begin{pmatrix}
0 & d_x & 0\\
d_x & 0 & 0\\
0 & 0 & 0
\end{pmatrix},
P=\begin{pmatrix}
0 & d_x & -\frac{u^3_x}{u^2}\\
d_x & 0 & 0\\
\frac{u^3_x}{u^2} & 0 & 0
\end{pmatrix},
P=\begin{pmatrix}
0 & d_x & \frac{u^3_x}{u^3u^1-u^2} \\
d_x & 0 & \frac{-u^3 u^3_x}{u^3u^1-u^2} \\
\frac{-u^3_x}{u^3 u^1-u^2}& \frac{u^3 u^3_x}{u^3u^1-u^2} & 0
\end{pmatrix},
\end{array}
\end{equation}

\begin{equation}\label{rank2_complex}
\begin{array}{c}
P=\begin{pmatrix}
d_x & 0 & 0\\
0 & d_x & 0\\
0 & 0 & 0
\end{pmatrix},
P=\begin{pmatrix}
d_x & 0 & 0\\
0 & d_x & -\frac{u^3_x}{u^2}\\
0 & \frac{u^3_x}{u^2} & 0
\end{pmatrix},
P=\begin{pmatrix}
d_x & 0 & \frac{-u^3 u^3_x}{u^3u^1-u^2}\\
0& d_x & \frac{u^3_x}{u^3u^1-u^2} \\
\frac{u^3 u^3_x}{u^3u^1-u^2} & \frac{-u^3_x}{u^3u^1-u^2}& 0
\end{pmatrix}.
\end{array}
\end{equation}
Furthermore, the canonical forms \eqref{rank2} and \eqref{rank2_complex} are equivalent under complex transformations.
\end{thm}

The proof of this theorem follows by straightforward computation, see Appendix A.

%%%%
%%%%
%%%%
%%%%

Let us briefly discuss a known example related to the theory of Hamiltonian systems.
\begin{example}
Given a Poisson structure $P$ of the form \eqref{Ham1}, Hamiltonian systems of hydrodynamic type are generated by Hamiltonians of the form $H=\int h({\bf u})dx$:
\begin{equation}\label{SysHam}
u^i_t=P^{ij}\frac{\delta H}{\delta u^j}.
\end{equation}
Such systems appear in a wide range of applications in hydrodynamics, chemical kinetics, the Whitham averaging method, the theory of Frobenius manifolds and so on, see the review papers \cite{DN, Tsarev} for further details and  references.

In three-component case, one well-known example is given by the equations of one-dimensional gas dynamics:
$$
v_t=-v v_x-\frac{p_{\rho}}{\rho} \rho_x-\frac{p_{s}}{\rho} s_x, \quad \rho_t=-(\rho v)_x,  \quad s_t=-v s_x,
$$
where $u^1=v$ is the gas velocity, $u^2=\rho$ is the mass density, $u^3=s$ is the entropy density, and $p=p(\rho,s)$ is the gas pressure. This system is Hamiltonian \cite{Mo} and the Hamiltonian operator related to this system is \cite{Ve}
$$
P=
\begin{pmatrix}
0 & -1 & 0\\
-1 & 0 & 0\\
0 & 0 & 0
\end{pmatrix}
\frac{d}{dx}
+
\begin{pmatrix}
0 & 0 & \dfrac{s_x}{\rho}\\
0 & 0 & 0\\
- \dfrac{s_x}{\rho} & 0 & 0
\end{pmatrix},
$$
with Hamiltonian density $h(v,\rho,s)=\frac{1}{2} \rho v^2 + f(\rho,s)$, where the function $f(\rho,s)$ is connected with the pressure $p(\rho,s)$ by $\rho f_{\rho}-f=p$.
One can easily see that up to a change of sign, this structure is equivalent to $\eqref{rank2}_2$. 
\end{example}

%%%%%
%%%%%
%%%%%

\section{Deformations of degenerate structures}\label{sect_deformations}
In this section we discuss deformations up to order 2 of the two-component Poisson structures we have classified, and investigate which of those deformations can be obtained by  Miura transformations. The Miura-group coincides with the semidirect product of the subgroup of diffeomorphisms (local change of coordinates) on the manifold $M$ and the subgroup of Miura-type transformations close to identity
\begin{equation}\label{miura_identity}
u^i \to u^i + \epsilon A^i_j(u) u^j_x +\epsilon^2 \left(B^i_j(u) u^j_{xx} +\frac{1}{2}C^i_{jk}(u) u^j_x u^k_x \right) +\dots,
\end{equation}
see \cite{DZ} for further details.
As we will see, the action of the subgroup of diffeomorphisms is not straightforward: it leads to several branches. Thus, for simplicity, we firstly discuss the action of the subgroup of Miura-type transformations close to identity (Section \ref{sect_miura_inf}). Then, in Section \ref{diffeo}, we analyse the action of local changes of coordinates.

Even though higher-order deformations can be obtained following the same procedure, the computations become much more complicated. Furthermore, we also analyse some examples of first-order deformations for three-component structures.

As defined in the introduction, a deformation of order $k$ of a $n$-component Poisson bivector $P_0$ is a formal series
$$
P_{\epsilon}=P_0 + \epsilon P_1 + \epsilon^2 P_2 + \ldots
$$
satisfying the condition $[P_{\epsilon}, P_{\epsilon}]=\mathcal{O}(\epsilon^k)$, where each coefficient $P_k$ has degree $k+1$, and is given by
$$
P_k =\sum_{s=0}^{k+1} A_s({\bf u}, {\bf u}_x, \ldots, {\bf u}_{k+1}) \frac{d^{k+1-s}}{dx^{k+1-s}}, \quad \mathrm{deg}(A_s)=s.
$$
The form of the operator $P_k$ depends on an increasing number of arbitrary functions of the coordinates $u^i$, $i=1, \ldots, n$.
Furthermore, these functions must be chosen in such a way that $P_k$ is skew-symmetric, namely $P_k^*=-P_k$.

In particular, the first two coefficients, $P_1$ and $P_2$, have the form
\begin{eqnarray*}
P^{ij}_1&=&A^{ij}({\bf u}) \frac{d^2}{dx^2}+ \sum_k B^{ij}_k({\bf u}) u^k_x \frac{d}{dx}+ \sum_k C^{ij}_k({\bf u}) u^k_{xx}+ \sum_{r\le k} D^{ij}_{rk}({\bf u}) u^r_x u^k_x,\\
P^{ij}_2&=&E^{ij}({\bf u}) \frac{d^3}{dx^3}+ \sum_k F^{ij}_k({\bf u}) u^k_x \frac{d^2}{dx^2}+ \left(\sum_k G^{ij}_k({\bf u}) u^k_{xx}+ \sum_{r\le k} H^{ij}_{rk}({\bf u}) u^r_x u^k_x\right) \frac{d}{dx}\\
&&+\sum_k L^{ij}_k({\bf u}) u^k_{xxx}+ \sum_{k,r} M^{ij}_{kr}({\bf u}) u^k_{xx} u^r_x \sum_{s\le r \le k} +N^{ij}_{srk}({\bf u}) u^s_x u^r_s u^k_x.
\end{eqnarray*}
This means that $P_1$ is defined by 
$$
n^2+n^3+n^3+n^2 \frac{n(n+1)}{2}=\frac{n^2 (n^2+5 n + 2)}{2}
$$
functions depending on the variables $u^1, \ldots, u^n$, while $P_2$ is given by
$$
\frac{n^2 (n^2+5 n + 2)}{2}+n^3+n^4+n^2\frac{n(n+1)(n+2)}{6}=\frac{n^2(n+2)(n^2+10n+3)}{6}
$$
functions in the variables $u^1, \ldots, u^n$.
Thus, one can see that the number of unknown functions is quite high already for a low number of components: for $n=2$ we have 104 unknowns, while for $n=3$ they are 432. Of course, imposing the skew-symmetry condition, this number falls.

\begin{rmk}
In order to simplify the computations, it is convenient to substitute the coefficients $D, H, N$ with $\tilde D, \tilde H, \tilde N$ such that
$$
\tilde{D}^{ij}_{rk}=\tilde{D}^{ij}_{kr}=\frac{1}{2}D^{ij}_{rk} \quad \mbox{if} \quad r< k, \quad \mbox{otherwise} \quad \tilde{D}^{ij}_{kk}=D^{ij}_{kk},
$$
$$
\tilde{H}^{ij}_{rk}=\tilde{H}^{ij}_{kr}=\frac{1}{2}H^{ij}_{rk} \quad \mbox{if} \quad r< k, \quad \mbox{otherwise} \quad \tilde{H}^{ij}_{kk}=H^{ij}_{kk},
$$
$$
\tilde{N}^{ij}_{srk}=\tilde{N}^{ij}_{ksr}=\tilde{N}^{ij}_{rks}=\tilde{N}^{ij}_{skr}=\tilde{N}^{ij}_{krs}=\tilde{N}^{ij}_{rsk}=\frac{1}{6} N^{ij}_{srk} \quad \mbox{if} \quad s< r< k,
$$
$$
\tilde{N}^{ij}_{rrs}=\tilde{N}^{ij}_{rsr}=\tilde{N}^{ij}_{srr}=\frac{1}{3} N^{ij}_{rrs} \quad \mbox{if} \quad  r < s,
$$
$$
\tilde{N}^{ij}_{rrs}=\tilde{N}^{ij}_{rsr}=\tilde{N}^{ij}_{srr}=\frac{1}{3} N^{ij}_{srr} \quad \mbox{if} \quad  r > s,
$$
$$
\tilde{N}^{ij}_{rrr}=N^{ij}_{rrr}.
$$
In this way, the summations involving these coefficients become
\begin{gather*}
\sum_{r\le k} D^{ij}_{rk}({\bf u}) u^r_x u^k_x=\sum_{r, k} \tilde{D}^{ij}_{rk}({\bf u}) u^r_x u^k_x,
\quad
\sum_{r\le k} H^{ij}_{rk}({\bf u}) u^r_x u^k_x=\sum_{r, k} \tilde{H}^{ij}_{rk}({\bf u}) u^r_x u^k_x,\\
\sum_{s\le r \le k} N^{ij}_{srk}({\bf u}) u^s_x u^r_s u^k_x=\sum_{s, r, k} \tilde{N}^{ij}_{srk}({\bf u}) u^s_x u^r_x u^k_x.
\end{gather*}
\end{rmk}
%%%%

\begin{lemma}\label{lemma_skew}
A first-order deformation is skew-symmetric if and only if the following conditions hold
\begin{subequations}
\begin{equation}
A^{ij}=-A^{ji},
\end{equation}
\begin{equation}
B^{ij}_k=-2\partial_k A^{ji}+B^{ji}_k,
\end{equation}
\begin{equation}
C^{ij}_k=-\partial_k A^{ji}+B^{ji}_k-C^{ji}_k,
\end{equation}
\begin{equation}
\tilde{D}^{ij}_{rk}=-\partial_r \partial_k A^{ji}+\frac{1}{2}\left(\partial_r B^{ji}_k+\partial_k B^{ji}_r\right)-\tilde{D}^{ji}_{rk}.
\end{equation}
\end{subequations}
Providing that the above conditions are satisfied, a second-order deformations is skew-symmetric if and only if the following conditions hold
\begin{subequations}
\begin{equation}
E^{ij}=E^{ji},
\end{equation}
\begin{equation}
F^{ij}_k=3 \partial_k E^{ji}-F^{ji}_k,
\end{equation}
\begin{equation}
G^{ij}_k=3 \partial_k E^{ji}-2 F^{ji}_k+ G^{ji}_k,
\end{equation}
\begin{equation}
\tilde{H}^{ij}_{rk}=3 \partial_r \partial_k E^{ji}-\partial_r F^{ji}_k-\partial_k F^{ji}_r+\tilde{H}^{ji}_{rk},
\end{equation}
\begin{equation}
L^{ij}_k=\partial_k E^{ji}-F^{ji}_k + G^{ji}_k - L^{ji}_k,
\end{equation}
\begin{equation}
M^{ij}_{rk}=3 \partial_r \partial_k E^{ji}-2 \partial_k F^{ji}_r-\partial_r F^{ji}_k+\partial_k G^{ji}_r+2\tilde{H}^{ji}_{rk}-M^{ji}_{rk},
\end{equation}
\begin{equation}
\begin{array}{c}
\tilde{N}^{ij}_{srk}=\partial_s \partial_r \partial_k E^{ji}-\frac{1}{3}\left(\partial_s \partial_r F^{ji}_k +\partial_r \partial_k F^{ji}_s + \partial_k \partial_s F^{ji}_r \right)\\
+\frac{1}{3}\left(\partial_s \tilde{H}^{ji}_{rk}+\partial_r \tilde{H}^{ji}_{ks}+\partial_k \tilde{H}^{ji}_{sr}\right)-\tilde{N}^{ji}_{srk}.
\end{array}
\end{equation}
\end{subequations}
\end{lemma}
A sketch of the proof can be found in Appendix B.
For instance, for $n=2$ the number of unknown functions falls to $12+30=42$.

\subsection{The action of infinitesimal Miura transformations}\label{sect_miura_inf}
Let us start with deformations of order 1. These deformations have to satisfy the Jacobi condition $[P_0,P_1]=0$. We want to eliminate deformations that can be obtained by an infinitesimal Miura transformation, that is, those that can be written as $\mathrm{Lie}_X P_0$, where $X$ is a suitable vector field of degree 1. In the non-degenerate case, it has been proved that all deformations of order 1 can be written in this way, but we will show that in the degenerate case this is not true.

Secondly, concerning deformations of order 2, namely
$$
P_{\epsilon}=P_0+\epsilon P_1 + \epsilon^2 P_2+\mathcal{O}(\epsilon^3),
$$
we have to consider the Jacobi condition
$
2[P_0,P_2]+[P_1,P_1]=0.
$
In our cases, first-order deformations $P_{\epsilon}$ cannot be reduced to $P_0$. In order to simplify the form of second-order deformations without changing lower order terms, we have to use infinitesimal Miura transformation like
\begin{equation}\label{Miura2}
\mathrm{Lie}_{Y} P_1+\mathrm{Lie}_Z P_0
\end{equation}
where $Z$ is an arbitrary vector field of degree 2 and $Y$ is a vector field of degree 1 which is a symmetry for $P_0$, namely $\mathrm{Lie}_{Y}P_0=0$.

To better understand this formula, let us consider the Lie series given by the vector field $\epsilon Y+\epsilon^2 Z$, we have
$$
\mathcal{L}_{\epsilon Y+\epsilon^2 Z} (P_{\epsilon})=P_0+ \epsilon (P_1+ \mathrm{Lie}_Y P_0)+\epsilon^2 \left(P_2+ \mathrm{Lie}_Y P_1+\frac{1}{2} \mathrm{Lie}^2_Y P_0+\mathrm{Lie}_Z P_0\right) + \mathcal{O}(\epsilon^3).
$$
Since $\mathrm{Lie}_Y P_0$ is assumed to vanish, we obtain exactly \eqref{Miura2}.

\subsubsection{Deformation results}

In two-component case, we have two different Poisson structures with degenerate metric (Theorem \ref{thm2cmpt}), one constant and one non-constant, which we call $P^{(1)}_0$ and $P^{(2)}_0$ respectively:
$$
P^{(1)}_0=
\begin{pmatrix}
d_x & 0 \\
0 & 0
\end{pmatrix},
\quad
P^{(2)}_0=
\begin{pmatrix}
d_x & -\frac{u^2_x}{u^1} \\
\frac{u^2_x}{u^1} & 0
\end{pmatrix}.
$$

\begin{thm}\label{thm_def1}
(1). All first-order deformations of $P^{(1)}_0$ can be reduced by infinitesimal Miura transformations to $P=P^{(1)}_0+\epsilon P_1 + \mathcal{O}(\epsilon^2)$ where
\begin{equation}\label{deg1def_case1}
P_1=
\begin{pmatrix}
0 & -p u^2_{xx} - q(u^2_x)^2 \\
p u^2_{xx} + q(u^2_x)^2 & r u^2_x d_x + \frac{1}{2} (r u^2_x)_x
\end{pmatrix},
\end{equation}
here $p, q, r$ are arbitrary functions of $u^2$.

\noindent
(2). All second-order deformations of $P^{(1)}_0$ can be reduced by infinitesimal Miura transformations to $P=P^{(1)}_0+\epsilon P_1 + \epsilon^2 P_2 + \mathcal{O}(\epsilon^3)$ where
$$
P_1=
\begin{pmatrix}
0 &-p u^2_{xx} - q(u^2_x)^2 \\
p u^2_{xx} +q(u^2_x)^2& 0
\end{pmatrix}
$$
and
\begin{equation}\label{deg2def1}
P_2=
\begin{pmatrix}
0 & 0 \\
0 & \alpha^{22}
\end{pmatrix}
\frac{d^3}{dx^3}
+\begin{pmatrix}
0 & 0 \\
0 & \beta^{22}
\end{pmatrix}
\frac{d^2}{dx^2}
+\begin{pmatrix}
0 & 0 \\
0 & \gamma^{22}
\end{pmatrix}
\frac{d}{dx}
+\begin{pmatrix}
0 & \eta^{12} \\
-\eta^{12} & \eta^{22}
\end{pmatrix},
\end{equation}
with
\begin{gather*}
\alpha^{22}=e, \quad \beta^{22}= \frac{3 e'}{2} u^2_x, \quad \gamma^{22} = g u^2_{xx} + h(u^2_x)^2,\\
\eta^{12}=(2p^2 u^1 - l) u^2_{xxx}  + pq u^1_x (u^2_x)^2+ p^2 u^1_x u^2_{xx}+\left( 2 u^1 (p q' + q^2) -n\right) (u^2_x)^3
\\
+\left(2 p u^1 (3q +p') - m \right)u^2_x u^2_{xx},
\end{gather*}
$$
\eta_2=\frac{1}{2} (g u^2_{xx}+h (u^2_x)^2)_x-\frac{1}{4}(e' u^2_{x})_{xx},
$$
where $p, q, e, g, h, l, m, n$ are arbitrary functions of $u^2$, and $'$ denotes the derivative with respect to $u^2$. Furthermore, it is always possible to reduce to zero one of the two functions $m$ or $n$.
\end{thm}

\begin{thm}\label{thm_def2}
(1). All first-order deformations of $P^{(2)}_0$ can be reduced by infinitesimal Miura transformations to $P=P^{(2)}_0+\epsilon P_1 + \mathcal{O}(\epsilon^2)$ where
\begin{equation}\label{deg1def2}
P_1=
\begin{pmatrix}
0 & 0 \\
0 & \frac{r}{(u^1)^3} u^2_x
\end{pmatrix}
\frac{d}{dx}+
\begin{pmatrix}
0 & -\frac{s}{(u^1)^3} (u^2_x)^2  \\
\frac{s}{(u^1)^3} (u^2_x)^2 & \frac{1}{2}\left(\frac{r}{(u^1)^3} u^2_x\right)_x
\end{pmatrix},
\end{equation}
here $r,s $ are arbitrary functions of $u^2$.

\noindent
(2). All second-order deformations of $P^{(2)}_0$ can be reduced by infinitesimal Miura transformations to $P=P^{(2)}_0+\epsilon P_1 + \epsilon^2 P_2 + \mathcal{O}(\epsilon^3)$ where $P_1$ is given by \eqref{deg1def2} and
\begin{equation}\label{deg2def2}
P_2=
\begin{pmatrix}
0 & 0\\
0 & \alpha^{22}
\end{pmatrix}
\frac{d^3}{dx^3}
+
\begin{pmatrix}
0 & 0\\
0 & \beta^{22}
\end{pmatrix}
\frac{d^2}{dx^2}
+
\begin{pmatrix}
0& \gamma^{12}\\
\gamma^{12}& \gamma^{22}
\end{pmatrix}
\frac{d}{dx}
+
\begin{pmatrix}
0 & \eta^{12}\\
\eta^{21} & \eta^{22}
\end{pmatrix},
\end{equation}
with
$$
\alpha^{22}=\frac{r^2}{2 (u^1)^4},\quad \beta^{22}=\frac{3 r r'}{2 (u^1)^4} u^2_x-\frac{3r^2}{(u^1)^5}u^1_x,
\quad
\gamma^{12}=\frac{19 s r}{6 (u^1)^5} \left(u^1 u^2_{xx}- u^1_x u^2_x \right),
$$
\begin{eqnarray*}
\gamma^{22}&=& \frac{15 r^2}{2 (u^1)^6} (u^1_x)^2 -\frac{2r^2}{(u^1)^5}u^1_{xx} -\frac{1}{(u^1)^5}\left(\frac{9r r'}{2}+ p \right) u^1_x u^2_x +\frac{p}{(u^1)^4} u^2_{xx},\\
\eta^{12}&=&
\frac{5 s r}{2 (u^1)^4} u^2_{xxx}
-\frac{5 sr }{2(u^1)^5} u^1_{xx} u^2_{x}
-\frac{32 sr }{3(u^1)^5} u^1_{x} u^2_{xx}
+\frac{3 s r' + s' r}{(u^1)^4}u^2_{x} u^2_{xx}
\\&&
+\frac{32 sr}{3 (u^1)^6} (u^1_{x})^2 u^2_{x}
-\frac{3 s r' + s' r}{(u^1)^5}u^1_{x} (u^2_{x})^2
-\frac{2 s^2}{(u^1)^5} (u^2_x)^3,
\\
\eta^{21}&=&
\frac{2 s r}{3 (u^1)^4} u^2_{xxx}
-\frac{2 sr }{3(u^1)^5} u^1_{xx} u^2_{x}
-\frac{31 sr }{6(u^1)^5} u^1_{x} u^2_{xx}
+\frac{13 s' r + s r'}{6 (u^1)^4}u^2_{x} u^2_{xx}
\\&&
+\frac{31 sr}{6 (u^1)^6} (u^1_{x})^2 u^2_{x}
-\frac{13 s' r + s r'}{6 (u^1)^5}u^1_{x} (u^2_{x})^2
+ \frac{2 s^2}{(u^1)^5} (u^2_x)^3,\\
\eta^{22}&=&
\frac{1}{2(u^1)^5} \left(\frac{3 r r'}{2}-5 p \right) u^1_{x} u^2_{xx}
-\frac{15 r^2}{2 (u^1)^7} (u^1_x)^3
-\frac{1}{2(u^1)^5} \left(\frac{5 r r'}{2}+p \right) u^1_{xx} u^2_{x}\\
&&
+\frac{1}{2 (u^1)^4} \left(p' -\frac{3}{2}\left( (r')^2+r r'' \right) \right) u^2_{x} u^2_{xx}
+\frac{5}{2 (u^1)^6} \left(\frac{3r r'}{2}+ p \right) (u^1_x)^2 u^2_x\\
&&
+\frac{1}{2 (u^1)^4} \left(p-\frac{r r'}{2} \right) u^2_{xxx}
-\frac{r^2}{2 (u^1)^5} u^1_{xxx}
-\frac{1}{4 (u^1)^4} \left( 3 r' r'' + r r'''\right) (u^2_x)^3\\
&&
+\frac{5 r^2}{(u^1)^6} u^1_x u^1_{xx}
-\frac{1}{2 (u^1)^5} \left(\frac{3}{2}\left(p'- (r')^2+r r''\right) \right) u^1_{x} (u^2_{x})^2,
\end{eqnarray*}
where $r, s, p$ are arbitrary functions of $u^2$ and $'$ denote the derivative with respect to $u^2$.
\end{thm}

The classification of three-component Poisson structures with degenerate metric is quite extensive (Theorem \ref{thm3cmpts}), so we have decided to study only some of them. Especially, we describe first-order deformations for the following operators:
$$
P_0^{(3)}=
\begin{pmatrix}
0 & u^3_x & 0\\
-u^3_x & 0 & 0\\
0 & 0 & 0
\end{pmatrix},
\quad
P_0^{(4)}=
\begin{pmatrix}
d_x & 0 & 0\\
0 & 0 & 0\\
0 & 0 & 0
\end{pmatrix},
\quad
P_0^{(5)}=
\begin{pmatrix}
d_x & 0 & 0\\
0 & d_x & 0\\
0 & 0 & 0
\end{pmatrix}.
$$

\begin{thm}\label{thm_def3}
All first-order deformations of $P^{(3)}_0$ can be reduced by infinitesimal Miura transformations to  $P=P^{(3)}_0+\epsilon P_1 + \mathcal{O}(\epsilon^2)$, where
\begin{equation}\label{3def0}
P_1=
\begin{pmatrix}
0& -\alpha^{21} & 0\\
\alpha^{21} & 0 & 0\\
0 & 0 & 0
\end{pmatrix}
\frac{d^2}{dx^2}
+
\begin{pmatrix}
\beta^{11} & \beta^{12} & \beta^{13}\\
\beta^{21} & \beta^{22} & \beta^{23}\\
\beta^{13} & \beta^{23} & 0
\end{pmatrix}
\frac{d}{dx}
+
\begin{pmatrix}
\gamma^{11} & \gamma^{12} & \gamma^{13}\\
\gamma^{21} & \gamma^{22} & \gamma^{23}\\
\gamma^{31} & \gamma^{32} & 0
\end{pmatrix}
\end{equation}
with
\begin{eqnarray*}
\alpha^{21}&=& a,\\
\beta^{11}&=& (2 \partial_2 a -b^{21}_2- \partial_2 s -\partial_2 r ) u^1_x + b^{11}_2 u^2_x,\\
\beta^{12}&=& (\partial_1 s - 2 \partial_1 a ) u^1_x + (b^{21}_2-2 \partial_2 a) u^2_x - 2 \partial_3 a u^3_x,\\
\beta^{13}&=& \frac{b^{21}_2+\partial_2 s +\partial_2 r}{2} u^3_x,\\
\beta^{21}&=& \partial_1 s u^1_x + b^{21}_2 u^2_x,\\
\beta^{22}&=& b^{22}_1 u^1_x + \partial_1 r u^2_x,\\
\beta^{23}&=& -\left(\partial_1 s +\frac{\partial_1 r}{2} \right) u^3_x,\\
\gamma^{11}&=&
\left( \partial_2 a -\frac{b^{21}_2-\partial_2 s -\partial_2 r}{2}\right) u^1_{xx}
+
\left(  \partial_1 \partial_2 a -\frac{ \partial_1 b^{21}_2-\partial_1 \partial_2 s -\partial_1 \partial_2 r}{2} \right) (u^1_x)^2
\\&&
+
\left( \partial_2 \partial_3 a -\frac{\partial_3 b^{21}_2-\partial_2 \partial_3 s -\partial_2 \partial_3 r}{2}  \right) u^1_x u^3_x
+
\frac{\partial_3 b^{11}_2}{2} u^2_x u^3_x
+
\frac{b^{11}_2}{2} u^2_{xx}
\\&&
+
\left( \partial_2^2 a - \frac{\partial_2 b^{21}_2 +\partial_2^2 r + \partial_2^2 s-\partial_1 b^{11}_2}{2} \right) u^1_x u^2_x
+
\frac{\partial_2 b^{11}_2}{2} (u^2_x)^2,
\\
\gamma^{12}&=&
\left(\frac{\partial_2^2 s + \partial_2^2 r + 3\partial_2 b^{21}_2-\partial_1 b^{11}_2}{4}-\frac{\partial_2^2 a}{2} \right) (u^2_x)^2
+\left(\partial_1 \partial_3 s - 2 \partial_1 \partial_3 a \right) u^1_x u^3_x
\\&&
+
\left(\frac{3\partial_1 b^{21}_2+3\partial_1\partial_2 s}{2} +\partial_1\partial_2 r -\partial_2^2 a \right) u^1_x u^2_x
+\left(\partial_3 b^{21}_2 -\partial_2\partial_3 a \right) u^2_x u^3_x
\\&&
+\left( \frac{3 \partial_1^2 s}{2}+\frac{\partial_2 b^{22}_1+\partial_1^2 r}{4} -\partial_1^2 a\right) (u^1_x)^2
+(b^{21}_2-\partial_2 a) u^2_{xx}-\partial_3 a u^3_{xx}
\\&&
+(\partial_1 s -\partial_1 a ) u^1_{xx}
-\partial_3^2 a  (u^3_x)^2,
\\
\gamma^{13}&=&
(\partial_2 s + \partial_2 r) u^3_{xx}+\frac{\partial_1 b^{21}_2-\partial_1 \partial_2 s}{2} u^1_x u^3_x
+\frac{\partial_3 b^{21}_2 +\partial_2 \partial_3 s +\partial_2 \partial_3 r }{2} (u^3_x)^2,
\\
\gamma^{21}&=&
 \left(\frac{\partial_2 b^{21}_2+\partial_1 b^{11}_2 -\partial_2^2 s -\partial_2^3 r}{4}-\frac{\partial_2^2 a}{2} \right) (u^2_x)^2
 -\left( \frac{\partial_2 b^{22}_1 +\partial_1^2 r}{4} +\frac{\partial_1^2 s}{2}\right) (u^1_x)^2
 \\&&
-\left(\frac{\partial_1 b^{21}_2 +\partial_1 \partial_2 s}{2} +\partial_1 \partial_2 r \right) u^1_x u^2_x
-\partial_2 \partial_3 a u^2_x u^3_x,
\\
\gamma^{22}&=&
\frac{b^{22}_1}{2}u^1_{xx}+\frac{\partial_1 r}{2} u^2_{xx}+\frac{\partial_3 b^{22}_1}{2}u^1_x u^3_x+\frac{ \partial_2 b^{22}_1+\partial_1^2 r}{2} u^1_x u^2_x
+\frac{\partial_1 \partial_3 r}{2}u^2_x u^3_x
\\&&
+\frac{\partial_1 b^{22}_1}{2} (u^1_x)^2+\frac{\partial_1 \partial_2 r}{2} (u^2_x)^2,
\\
\gamma^{23}&=&
\frac{ \partial_1 b^{21}_2-\partial_1 \partial_2 s }{2}u^2_x u^3_x
-\left( \partial_1 \partial_3 s + \frac{\partial_1\partial_3 r}{2}\right) (u^3_x)^2
-(\partial_1 s + \partial_1 r)u^3_{xx},
\\
\gamma^{31}&=&
\frac{b^{21}_2 -\partial_2 s -\partial_2 r}{2} u^3_{xx}+\left( \partial_1 \partial_2 s + \frac{\partial_1 \partial_2 r}{2}\right) u^1_x u^3_x
+\frac{\partial_2 b^{21}_2 + \partial_2^2 s + \partial_2^2 r}{2} u^2_x u^3_x,
\\
\gamma^{32}&=&
\frac{\partial_1 r}{2} u^3_{xx}-\left( \partial_1^2 s + \frac{\partial_1^2 r}{2}\right) u^1_x u^3_x
-\frac{\partial_1 b^{21}_2 +\partial_1 \partial_2 s +\partial_1 \partial_2 r}{2} u^2_x u^3_x,
\end{eqnarray*}
where $a, r, s, b^{11}_2, b^{21}_2, b^{22}_1$are arbitrary functions of $u^1, u^2, u^3$, and $\partial_k$ means partial derivative with respect to $u^k$, for $k=1,2,3$.
\end{thm}

\begin{thm}\label{thm_def4}
All first-order deformations of $P^{(4)}_0$ can be reduced by infinitesimal Miura transformations to  $P=P^{(4)}_0+\epsilon P_1 + \mathcal{O}(\epsilon^2)$, where
\begin{equation}\label{3def1}
P_1=
\begin{pmatrix}
0& 0 & 0\\
0 & 0 & -\alpha^{32}\\
0 & \alpha^{32} & 0
\end{pmatrix}
\frac{d^2}{dx^2}
+
\begin{pmatrix}
0 & 0 & 0\\
0 & \beta^{22} & \beta^{23}\\
0 & \beta^{32}  & \beta^{33} 
\end{pmatrix}
\frac{d}{dx}
+
\begin{pmatrix}
0 & -\gamma^{21} & -\gamma^{31}\\
\gamma^{21} & \gamma^{22} & \gamma^{23}\\
\gamma^{31} & \gamma^{32} & \gamma^{33}
\end{pmatrix}
\end{equation}
with
$$
\alpha^{32}=a, \quad
\beta^{ij}= b^{ij}_2 u^2_x+b^{ij}_3 u^3_x \quad (i\ge j), \quad \beta^{23}=\beta^{32}-2 a_x,
$$
$$
\gamma^{ij}=  c^{ij}_2 u^2_{xx}+c^{ij}_3 u^3_{xx} +e^{ij}_{22} (u^2_x)^2 + e^{ij}_{23} u^2_x u^3_x+ e^{ij}_{33} (u^3_x)^2  \quad (i> j),
$$
$$
\gamma^{23}=\beta^{32}_x - a_{xx}- \gamma^{32}, \quad
\gamma^{ii}=\frac{1}{2} \beta^{ii}_x,
$$
where $a$, $b^{rs}_k$, $c^{ij}_k$, $e^{ij}_{mk}$ (for $r\ge s$ and $k\ge m$ and $i>j$, where $i,j=1,2,3$ and $r,s,m,k=2,3$) are arbitrary functions of $u^2, u^3$.
\end{thm}

\begin{thm}\label{thm_def5}
All first-order deformations of $P^{(5)}_0$ can be reduced by infinitesimal Miura transformations to  $P=P^{(5)}_0+\epsilon P_1 + \mathcal{O}(\epsilon^2)$, where
\begin{equation}\label{3def2}
P_1=
\begin{pmatrix}
0 & 0 & 0\\
0 & 0 & 0\\
0 & 0 & \beta^{33}
\end{pmatrix}
\frac{d}{dx}
+
\begin{pmatrix}
0 & -\gamma^{21} & -\gamma^{31}\\
\gamma^{21} & 0 & -\gamma^{32}\\
\gamma^{31} & \gamma^{32} & \gamma^{33}
\end{pmatrix}
\end{equation}
with
$$
\beta^{33}= b u^3_x,  \quad \gamma^{33} = \frac{1}{2} \left(b u^3_x \right)_x, \quad \gamma^{ij} = e^{ij} (u^3_x)^2 + c^{ij} u^3_{xx} \quad (i>j), \\
$$
where $b, c^{ij}, e^{ij}$, for $i>j$, are arbitrary functions of $u^3$. Furthermore, it is always possible to reduce to zero one of the functions $e^{21}$ or $c^{21}$.
\end{thm}

The proofs of the above theorems are given in Appendix C.

\subsection{Local change of coordinates}\label{diffeo}
The classification provided in the previous section has been obtained using infinitesimal Miura transformations. As we pointed out above, the whole Miura group contains also local changes of coordinates which preserve the dispersionless limit of our structures.

\subsubsection{Two-component case}
Let us consider deformations of the structure $P^{(1)}_0$, Theorem \ref{thm_def1}. Local changes of coordinates which preserve the form of the dispersionsless term $P^{(1)}_0(u)$ are of the form $u^1=v^1 + \omega_1(v^2), u^2=\omega_2(v^2)$.
Let us apply this transformation to the bivector $P_1$ given by \eqref{deg1def_case1}, using the transformation rule $P(v)=J P(u) J^t$, where ${}^t$ means the transpose and $J^i_j=\frac{\partial v^i}{\partial u^j}$. We have
$$
J=
\begin{pmatrix}
1 & -\frac{\omega_1'}{\omega_2'}\\
0 & \frac{1}{\omega_2'}
\end{pmatrix},
$$
where prime denotes the derivative with respect to $v^2$. Looking at the coefficient of $d_x$ in \eqref{deg1def_case1}, it transforms as
$$
\begin{pmatrix}
0 & 0 \\
0 & r(u^2) u^2_x
\end{pmatrix}
\quad \mapsto \quad
\begin{pmatrix}
- \frac{(\omega_1')^2}{\omega_2'} r(\omega_2) v^2_x & \frac{\omega_1'}{\omega_2'} r(\omega_2) v^2_x  \\
\frac{\omega_1'}{\omega_2'}  r(\omega_2) v^2_x  & \frac{1}{\omega_2'}  r(\omega_2) v^2_x 
\end{pmatrix}.
$$
In the general case where $r\neq0$, this transformation suggests to set $\omega_1'=0$, otherwise we would have a new arbitrary function in the coefficient of $d_x$ in $P_1$. Without any loss of generality, at this stage we can consider $\omega_1=0$. Looking at the whole bivector $P^1$, by straightforward computation, we get the following rule for the arbitrary functions $r, p, q$:
$$
r(u^2) \mapsto \frac{r(\omega_2)}{\omega_2'}, \quad
p(u^2) \mapsto p(\omega_2), \quad
q(u^2) \mapsto p(\omega_2)\frac{\omega_2''}{\omega_2'} + q(\omega_2) \omega_2',
$$
(if $r=0$, the action of the local change is still the same, namely, the function $\omega_1$ is not involved in the transformation of $p$ and $q$).
Thus, with a suitable choice of $\omega_2$, one can eliminate the function $q$.

Looking at the deformations of order two \eqref{deg2def1}, since $r=0$, we still have the freedom of one arbitrary function due to $\omega_1$. Suppose we have used $\omega_2$ to simplify $p$ or $q$. Then, the coefficient of $d_x^3$ transforms as
$$
\begin{pmatrix}
0 & 0 \\
0 & e(u^2)
\end{pmatrix}
\quad \mapsto \quad
\begin{pmatrix}
- (\omega_1')^2 e(v^2) &\omega_1' e(v^2)  \\
\omega_1' e(v^2)  & e(v^2) 
\end{pmatrix}.
$$
Once again, this means that $\omega_1'=0$, otherwise we would have an extra function.
Summarising, up to diffeomorphisms, we are able to simplify at most only one arbitrary function in the first and second deformation of $P^{(1)}_0$.

Considering $P^{(2)}_0$, a generic change of coordinates which preserves its form is given by $u^1=v^1, u^2=\omega(v^2)$.
Here, looking at first-order deformations, Theorem \ref{thm_def2}, the two arbitrary functions $r$ and $s$ appearing in $P_1$ transform as
$$
s(u^2) \mapsto s(\omega_2) \omega_2', \quad r(u^2) \mapsto \frac{r(\omega_2)}{\omega_2'}.
$$
Therefore, also in this case we can simplify at most one single function.

\subsubsection{Three-component case}

Although the analysis of three-component case can be performed in the same way, computations become much more complicated. Therefore, it is not always possible to provide a complete description of the action of local change of coordinates on the structures we studied.
In this subsection, we are going to describe the action of the group of diffeomorphisms on second-order deformations of $P_0^{(5)}$, since this is the only case where we can provide a detailed analysis.

Up to infinitesimal Miura transformations, a first order deformation of $P_0^{(5)}$ reduces to the one described in Theorem \ref{thm_def5}. Diffeomorphisms which preserve the form of $P_0^{(5)}$ are
$$
u^1=v^1 \cos{\kappa}+ v^2\sin{\kappa} + \varphi_1(v^3), \quad u^2=v^1 \sin{\kappa}-v^2\cos{\kappa} + \varphi_1(v^3), \quad u^3=\varphi_3(v^3),
$$
Without any loss of generality, we can set $\kappa=0$.
The coefficient of $d_x$ in \eqref{3def2} transforms as
$$
\begin{pmatrix}
0 & 0 & 0\\
0 & 0 & 0\\
0 & 0 & b(u^3)
\end{pmatrix}
\quad \mapsto \quad
\begin{pmatrix}
\frac{(\varphi_1')^2 b(\varphi_3) v^3_x}{\varphi_3'} & \frac{\varphi_1' \varphi_2' b(\varphi_3) v^3_x}{\varphi_3'}  & \frac{-\varphi_1' b(\varphi_3) v^3_x}{\varphi_3'} \\
\frac{\varphi_1' \varphi_2' b(\varphi_3) v^3_x}{\varphi_3'}  & \frac{(\varphi_2')^2 b(\varphi_3) v^3_x}{\varphi_3'}  & \frac{-\varphi_2' b(\varphi_3) v^3_x}{\varphi_3'}\\
\frac{-\varphi_1' b(\varphi_3) v^3_x}{\varphi_3'}  & \frac{-\varphi_2' b(\varphi_3) v^3_x}{\varphi_3'}  & \frac{ b(\varphi_3) v^3_x}{\varphi_3'} 
\end{pmatrix},
$$
here $'$ denote the derivative with respect to $v^3$. Therefore, when $b\neq 0$, we have to impose $\varphi_i'=0$, for $i=1,2$, otherwise two new functions would appear in the coefficient of $d_x$. Setting $\varphi_i=\xi_i$, where $\xi_i=const$,  $i=1,2$, the functions appearing in \eqref{3def2} transform as
$$
b \mapsto \frac{b}{\varphi_3'},
\; \;
e^{21}\mapsto e^{21}(\varphi_3')^2+c^{21}\varphi_3'',
\; \;
c^{21}\mapsto c^{21}\varphi_3',
\; \;
e^{3j}\mapsto \frac{e^{3j}(\varphi_3')^2+c^{3j}\varphi_3''}{\varphi_3'},
\; \;
c^{3j}\mapsto c^{3j},
$$
for $j=1,2$ (here $e^{ij}, c^{ij}$ on the left hans side, with respect to the arrow, depend on $u^3$, while on the right hand side they depend on $\varphi_3(v^3)$).
Thus, in the most general case, namely $b\neq 0$, local change of coordinates allows to reduce by one the number of arbitrary functions appearing in the deformation. For instance, we can choose to reduce $b$ to 1.
Let us recall that infinitesimal Miura transformations allow to reduce to 0 one function between $e^{21}$ and $c^{21}$.
Thus, we have the following
\begin{thm}
Up to Miura transformations, a generic second-order deformation of $P_0^{(5)}$ depends on 5 functions of $u^3$.
\end{thm}

At this point, one could ask: if $b=0$, how does the group of diffeomorphism act on the structure?
Although this is a reasonable question, a deeper analysis of this case does not provide any further information about the general form of the deformation we are studying. However, it is remarkable that under this strong assumption ($b=0$), we still have the freedom of all three arbitrary functions $\varphi_1, \varphi_2, \varphi_3$. Let us discuss this sub-case. Clearly, the number of arbitrary functions appearing in the deformation is already reduced by one, since $b$ is assumed to be zero. The functions $e^{ij}, c^{ij}$ transform as
$$
e^{21}\mapsto
e^{21}(\varphi_3')^2+c^{21}\varphi_3''
-\frac{\varphi_2'(e^{31}(\varphi_3')^2+c^{31}\varphi_3'')-\varphi_1'(e^{32}(\varphi_3')^2+c^{32}\varphi_3'')}{\varphi_3'},
$$
$$
c^{21}\mapsto
c^{21}\varphi_3'-c^{31}\varphi_2'+c^{32}\varphi_1',
\quad
e^{3j}\mapsto
\frac{e^{3j}(\varphi_3')^2+c^{3j}\varphi_3''}{\varphi_3'},
\quad
c^{3j}\mapsto
c^{3j}
$$
for $j=1,2$. Let us assume for simplicity that all $e^{ij}, c^{ij}$ are non-zero (otherwise, we should discuss case by case). Therefore, both $e^{21}$ and $c^{21}$ can be brought to $0$, using $\varphi_1, \varphi_2$. Finally, the freedom of $\varphi_3$ allows to simplify another functions between $e^{31}, e^{32}, c^{31}$ and $c^{32}$.

\begin{coro}
Let $b=0$ in \eqref{3def2}. Up to Miura transformations, second-order deformations of $P_0^{(5)}$ depend on 3 functions of $u^3$.
\end{coro}

Changes of local coordinates which preserve the form of the undeformed Poisson structure $P_0^{(3)}$ and $P_0^{(4)}$ are quite easy to compute. For $P_0^{(3)}$ these transformations are given by
$$
u^1=\varphi_1(v^1,v^2,v^3), \quad u^2=\varphi_2(v^1,v^2,v^3), \quad u^3=\varphi_3(v^3),
$$
with the constraint
$$
\partial_1 \varphi_1 \partial_2  \varphi_2 - \partial_2  \varphi_1 \partial_1  \varphi_2=\partial_3\varphi_3, \quad \partial_i=\frac{\partial}{\partial v^i},
$$
while for $P_0^{(4)}$ we have
$$
u^1=v^1+\varphi_1(v^2,v^3), \quad u^2=\varphi_2(v^2,v^3), \quad u^3=\varphi_3(v^2,v^3).
$$
Unfortunately, the action of these transformations on the respective deformed structures are very cumbersome, and we are not going to describe it.

Summarising, as we have seen, the action of the subgroup of diffeomorphisms leads to several branches for each case, depending wherever the functional parameters are constant, zero or arbitrary.
Furthermore, the number of additional arbitrary functions we can use, due to these transformations, is always lower than the number of functional parameters appearing in the deformations. This implies that, in each cases we have studied, we cannot reduce the deformation to its dispersionless term, and therefore the deformation is not trivial.

%%%%%%%%%%
%%%%%%%%%%

\section{Concluding remarks}
This paper is the first step towards the deformation theory for Poisson structures of hydrodynamic type with degenerate metric. Besides the complete list of two- and three-component Hamiltonian operators with degenerate metric, our main contributions include the proof that in two-component case, first- and second-order deformations are not trivial, as well as examples of non-trivial first-order deformations for some three-component structures. This implies that the second cohomology group for such structures is not trivial, contrary to what happens in the non-degenerate case.

Our results suggest the following
\begin{conj}
The $k$-order deformations of two-component Poisson structures with degenerate metric are characterised by functions depending on the single variable $u^2$.
\end{conj}
Unfortunately, the number of unknowns in this problem grows rapidly with the increase of the order of deformations, and computations become more and more complicated. Thus, it seems necessary to find a different approach in order to prove the conjecture. Moreover, there seems to be no rule which provides the number of arbitrary functions on which the deformations depend.

Furthermore, a deeper analysis of the three-component case would be an important step to better understand what happens in a more general contest, in order to generalise our results:
\begin{conj}
If the matrix $g$ which defines a $n$-component Poisson structure $P$ of the form \eqref{Ham1} has rank  $m<n$, the deformations of $P$ are characterised by arbitrary functions depending on the set of variables $(u^{m+1}, \ldots, u^n)$.
\end{conj}

\section*{Acknowledgments}
\addcontentsline{toc}{section}{Acknowledgments}
I would like to express my deep gratitude to Eugene Ferapontov and Paolo Lorenzoni for numerous suggestions. I also wish to thank Alexey Bolsinov for useful discussions.

%%%%%%%%%%
%%%%%%%%%%
%%%%%%%%%%
%%%%%%%%%%

%\appendix

\addcontentsline{toc}{section}{Appendix A. Proof of Theorem \ref{thm3cmpts}}
\section*{Appendix A. Proof of Theorem \ref{thm3cmpts}}

Here we give the proof of Theorem \ref{thm3cmpts}. Let us consider separately each case depending on the rank of the metric.

\vspace{5pt}
\noindent
{\bf Rank 0}.

\noindent
According to the result of Grinberg \cite{Grinberg}, if $b^{ij}_k$ are not identically zero, they reduce to constant form $b^{12}_{3}=-b^{21}_3=1$ and the remaining $b^{ij}_k=0$. Thus, the operator reads
$$
P=
\begin{pmatrix}
0 & u^3_x & 0\\
-u^3_x & 0 & 0\\
0 & 0 & 0
\end{pmatrix}.
$$
As noticed by Bogoyavlenskij \cite{B1}, the coefficients $b^{ij}_k$ in this case define the Heisenberg nilpotent Lie Algebra $N_3$: let us consider the basis of coordinate 1-forms $e_1, e_2, e_3$ on the cotangent spaces $T^{*}_u (\mathbb{R}^3)$, then $[e_1,e_2]=e_3$, $[e_1,e_3]=[e_2,e_3]=0$.

%%%%%
%%%%%
%%%%%

\vspace{5pt}
\noindent
{\bf Rank 1}.

\noindent
In the case $\mathrm{rank}(g^{ij})=1$, there exists a coordinate system where the metric assumes the canonical form
$$
g^{ij}=
\begin{pmatrix}
1 & 0 & 0\\
0 & 0 & 0\\
0 & 0 & 0
\end{pmatrix}.
$$
Let us point put that the (restricted) admissible transformations in this case are
$$
u^1=\tilde{u}^1+F_1(\tilde{u}^2,\tilde{u}^3), u^2=F_2(\tilde{u}^2,\tilde{u}^3), u^3=F_3(\tilde{u}^2,\tilde{u}^3),
$$
this means that we always can apply a permutation of $u^2,u^3$ without any problem, since the symbols $b^{ij}_{k}$ in this case transform as $(2,1)$-tensor.

Let us solve Grinberg's conditions \eqref{1D_cond}. We already know that $b^{ii}_k=0$, and the unknowns are $b^{ij}_k$ for $i<j$. The algebraic conditions given by \eqref{G3} imply $b^{12}_1=b^{13}_1=b^{23}_1=0$, while the algebraic relations given by \eqref{G4} imply $b^{23}_2=b^{23}_3=0$.
The remaining unknowns $b^{12}_2, b^{12}_3, b^{13}_2, b^{13}_3$ have to satisfy differential equations given by \eqref{G4} and \eqref{G5}. Let us call
$$
b^{12}_2=\mu, \quad b^{12}_3=\nu, \quad b^{13}_2=\phi, \quad b^{13}_3=\eta.
$$
In what follows the subscript $i$ means derivative with respect to $u^i$. Differential conditions given by \eqref{G4} read
\begin{equation}\label{rk1_eq1}
\mu_1=\mu^2+\nu \phi, \quad \nu_1=\nu (\mu+ \eta), \quad \phi_1=\phi (\mu+ \eta),  \quad \eta_1=\eta^2+\nu \phi,
\end{equation}
while the conditions given by \eqref{G5} are
\begin{equation}\label{rk1_eq2}
\phi_1 \mu=\phi \mu_1, \quad \eta_1 \nu=\eta \nu_1, \quad \phi_1 \nu +\eta_1 \mu=\mu_1 \eta + \nu_1 \phi.
\end{equation}
Using \eqref{rk1_eq1}, one can easily see that conditions \eqref{rk1_eq2} become algebraic:
\begin{equation}\label{rk1_eq3}
\phi(\eta \mu - \nu \phi)=0, \quad \nu(\eta \mu - \nu \phi)=0, \quad (\eta-\mu)(\eta \mu - \nu \phi)=0.
\end{equation}
Solving this algebraic system we get two different solutions:
\begin{gather}
\phi=0, \quad \nu=0, \quad \eta=\mu \label{sol3},\\
\phi\neq 0, \quad \nu=\dfrac{\eta \mu}{\phi} \label{sol4}.
\end{gather}

%%%%%%%%
%%%%%%%%

Before solving system \eqref{rk1_eq1}, let us point out that, since the change of coordinates $\tilde{u}^1=u^1,\tilde{u}^2=u^3,\tilde{u}^3=u^2$ transforms the operator
$$
P^{ij}=
\begin{pmatrix}
1 & 0 & 0\\
0 & 0 & 0\\
0 & 0 & 0
\end{pmatrix}\frac{d}{dx}+
\begin{pmatrix}
0 & \mu u^2_x +\nu u^3_x &  \phi u^2_x+\eta u^3_x\\
-\mu u^2_x -\nu u^3_x & 0 & 0\\
-\phi u^2_x-\eta u^3_x & 0 & 0
\end{pmatrix}
$$
to the form
$$
\tilde{P}^{ij}=
\begin{pmatrix}
1 & 0 & 0\\
0 & 0 & 0\\
0 & 0 & 0
\end{pmatrix}\frac{d}{dx}+
\begin{pmatrix}
0 &  \tilde{\phi} \tilde{u}^3_x+\tilde{\eta} \tilde{u}^2_x & \tilde{\mu} \tilde{u}^3_x +\tilde{\nu} \tilde{u}^2_x\\
-\tilde{\phi} \tilde{u}^3_x-\tilde{\eta} \tilde{u}^2_x& 0 & 0\\
-\tilde{\mu} \tilde{u}^3_x -\tilde{\nu} \tilde{u}^2_x& 0 & 0
\end{pmatrix},
$$
we can exchange the coefficients $\mu$, $\eta$ and $\nu$, $\phi$.

\noindent
{\bf Solution \eqref{sol3}.} If $\phi=\nu=0$ and $\eta=\mu$, conditions \eqref{rk1_eq1} lead to $\mu_1=\mu^2$. Thus $\phi=\nu=0, \eta=\mu$ and
$$
\mu=0, \quad \mbox{or} \quad \mu=\frac{1}{F-u^1},
$$
where $F=F(u^2,u^3)$ is an arbitrary function.
The case $\mu=0$ leads to the constant operator $\eqref{rank1}_1$,
$$
P^{ij}=
\begin{pmatrix}
d_x & 0 & 0\\
0 & 0 & 0\\
0 & 0 & 0
\end{pmatrix}.
$$
Otherwise, if $\mu=\frac{1}{F-u^1}$, applaying the change of coordinates $u^1=\tilde{u}^1+F$ we get $\mu=\eta=-\frac{1}{\tilde{u}^1}$. Thus, the operator can be brought to $\eqref{rank1}_4$,
$$
P^{ij}=
\begin{pmatrix}
d_x & -\frac{u^2_x}{u^1} & -\frac{u^3_x}{u^1}\\
\frac{u^2_x}{u^1} & 0 & 0\\
\frac{u^3_x}{u^1} & 0 & 0
\end{pmatrix}.
$$

\noindent
{\bf Solution \eqref{sol4}.} Assuming $\phi$ non-zero and $\nu=\frac{\eta \mu}{\phi}$, conditions \eqref{rk1_eq1} read
$$
\mu_1=\mu(\mu+\eta), \quad \phi_1=\phi (\mu+ \eta), \quad \eta_1=\eta(\mu +\eta),
$$
since the fourth one is fulfilled. By straightforward computation, the solutions of \eqref{rk1_eq1} are
\begin{equation}\label{rank1_sol1}
\mu=F, \quad \nu=\frac{-F^2}{R}, \quad \phi=R, \quad \eta=-F,
\end{equation}
and
\begin{equation}\label{rank1_sol2}
\mu=\frac{S}{F-(S+1)u^1},  \nu=\frac{-S}{R(F-(S+1)u^1)},  \phi=\frac{-R}{F-(S+1)u^1},  \eta=\frac{1}{F-(S+1)u^1},
\end{equation}
for arbitrary functions $F=F(u^2,u^3)$, $S=S(u^2,u^3)$ and $R=R(u^2,u^3)\neq0$.
Here we have to consider different cases.

\noindent
{\bf Case 1}. Let us assume $F=0$ in \eqref{rank1_sol1}. Choosing the transformation $u^1=\tilde{u}^1$, $u^2=\tilde{u}^2$, $u^3=W(\tilde{u}^2,\tilde{u}^3)$, we get
$$
\tilde{\phi}=\frac{R(\tilde{u}^2,W)}{W_3}.
$$
Thus we can always choose $W$ such that $\mu$ is reduced to $1$, obtaining (after interchanging $u^2$, $u^3$) the operator $\eqref{rank1}_2$
$$
P^{ij}=
\begin{pmatrix}
d_x & u^3_x& 0\\
-u^3_x& 0 & 0\\
0& 0 & 0
\end{pmatrix}.
$$

\noindent
{\bf Case 2}.
Otherwise, if $F$ is not zero in \eqref{rank1_sol1}, a transformation of the form $u^1=\tilde{u}^1$, $u^2=W(\tilde{u}^2,\tilde{u}^3)$, $u^3=W(\tilde{u}^2,\tilde{u}^3)$ implies
\begin{gather*}
\tilde{\mu}=\frac{(W_3 F - V_3 R)(V_2 R - W_2 F)}{(V_2 W_3 - W_2 V_3) R}, \quad
\tilde{\nu}=-\frac{(W_3 F - V_3 R)^2}{(V_2 W_3 - W_2 V_3) R},\\
\tilde{\phi}=\frac{(V_2 R - W_2 F)^2}{(V_2 W_3 - W_2 V_3) R}, \quad
\tilde{\eta}=-\frac{(W_3 F - V_3 R)(V_2 R - W_2 F)}{(V_2 W_3 - W_2 V_3) R}.
\end{gather*}
Thus, choosing $V, W$ such that $W_3 F - V_3 R=0$ and $V_2 R - W_2 F=W_3$, we obtain $\tilde{\phi}=1$, which leads to $\eqref{rank1}_2$.

Let us remark that we cannot choose both $W_3 F - V_3 R=0$ and $V_2 R - W_2 F=0$, otherwise we would have $V_2 W_3 - W_2 V_3=0$.

\noindent
{\bf Case 3}. Let us assume $S=0$ in \eqref{rank1_sol2}.  The change of variables $u^1=\tilde{u}^1+F$ allows to reduce $F$ to 0. The transformation
$
u^1=\tilde{u^1}, u^2=V (\tilde{u}^2), u^3=W (\tilde{u}^2,\tilde{u}^3)
$
preserves $\eta$ and transforms $\phi$ into
$$
\tilde{\phi}=-\frac{V_2 R + W_2}{W_3 \tilde{u}^1}.
$$
Thus, we can choose $V, W$ such that $\tilde{\phi}=0$, obtaining $\eqref{rank1}_3$,
$$
P^{ij}=
\begin{pmatrix}
d_x & 0& -\frac{u^3_x}{u^1}\\
0& 0 & 0\\
\frac{u^3_x}{u^1}& 0 & 0
\end{pmatrix}.
$$

\noindent
{\bf Case 4}. If $S= -1$, relabelling $F=-\frac{1}{Q}$ and $R=-\frac{T}{Q}$ we get
$$
\mu=Q, \quad \nu= -\frac{Q^2}{T}, \quad \phi=T, \quad \eta=-Q,
$$
which is the same as Case 2. 

\noindent
{\bf Case 5}. If $S\neq 0,-1$, choosing the transformation $u^1=\tilde{u}^1+\frac{F}{S+1}$ we can reduce $F$ to 0.
By tensorial calculus, one can see that a change of variables of the form
$$
u^1=\tilde{u}^1, \quad u^2=V (\tilde{u}^2,\tilde{u}^3), \quad u^3=W (\tilde{u}^2,\tilde{u}^3),
$$
transforms the coefficients $\mu, \nu, \phi, \eta$ as
\begin{gather*}
\tilde\mu=-\frac{(W_3 S+V_3 R)(W_2-V_2 R)}{(W_2 V_3-V_2 W_3)R (S+1) \tilde{u}^1}, \quad
\tilde\nu=-\frac{(W_3 S+V_3 R)(W_3-V_3 R)}{(W_2 V_3-V_2 W_3)R (S+1) \tilde{u}^1},\\
\tilde\phi=\frac{(W_2 S+V_2 R)(W_2-V_2 R)}{(W_2 V_3-V_2 W_3)R (S+1) \tilde{u}^1}, \quad
\tilde\eta=\frac{(W_2 S+V_2 R)(W_3-V_3 R)}{(W_2 V_3-V_2 W_3)R (S+1) \tilde{u}^1}.
\end{gather*}
If we choose the functions $V$ and $W$ such that satisfy
$$
V_3 R+W_3 S=0, \quad W_2-V_2 R=0,
$$
we obtain $\tilde\mu= \tilde\nu=\tilde\phi=0$ and $\eta=-\frac{1}{\tilde{u}^1}$, which leads to the operator $\eqref{rank1}_3$.

Finally, by straightforward computation it follows that these four canonical forms are not equivalent up to admissible changes of coordinates.

%%%%%
%%%%%
%%%%%

\vspace{5pt}
\noindent
{\bf Rank 2}.

\noindent
In this case, there always exists a coordinate system where the metric assumes one of the two canonical forms
\begin{equation}\label{metrics_rank2}
\begin{array}{cc}
\begin{pmatrix}
0 & 1 & 0\\
1 & 0 & 0\\
0 & 0 & 0
\end{pmatrix},
\quad & \quad
\begin{pmatrix}
1 & 0 & 0\\
0 & 1 & 0\\
0 & 0 & 0
\end{pmatrix}.
\\[10pt]
\mbox{\footnotesize \it Case (a)} \quad & \quad \mbox{\footnotesize \it Case (b)}
\end{array}
\end{equation}

\noindent
{\bf Case (a)}. Let as assume that the metric is
$$
g^{ij}=
\begin{pmatrix}
0 & 1 & 0\\
1 & 0 & 0\\
0 & 0 & 0
\end{pmatrix}.
$$
Conditions \eqref{G1}--\eqref{G3} imply that all the coefficients $b^{ij}_k$ must be zero except
$$
b^{12}_3=-b^{21}_3=\mu, \quad b^{13}_3=-b^{31}_3=\nu, \quad b^{23}_3=-b^{32}_3=\phi.
$$
Relation \eqref{G4} reads
\begin{equation}\label{rk2_eq1}
\mu_1=\mu \phi, \quad \nu_1=\nu \phi, \quad \phi_1=\phi^2, \quad
\mu_2=\mu \nu, \quad \nu_2=\nu^2, \quad \phi_2=\nu \phi,
\end{equation}
while \eqref{G5} leads to
$$
\nu_1 \phi=\phi_1 \nu, \quad \nu_2 \phi=\phi_2 \nu, \quad \phi_2 \mu +\nu_1 \mu=\mu_1 \nu+\mu_2 \phi.
$$
One can easily see that these last three equations are fulfilled if conditions given by \eqref{rk2_eq1} hold.

In order to solve this system, since we have $\phi_1=\phi^2$ and $\nu_2=\nu^2$, we should consider four different cases:
$$
\nu=\phi=0, \quad \nu=0, \phi\neq 0, \quad \nu\neq 0, \phi=0, \quad \nu\neq 0, \phi \neq 0.
$$
However, we can consider a permutation of $u^1, u^2$, which has no effect on the metric. Indeed, the operator
$$
P^{ij}=
\begin{pmatrix}
0 & 1 & 0\\
1 & 0 & 0\\
0 & 0 & 0
\end{pmatrix}\frac{d}{dx}+
\begin{pmatrix}
0 & \mu u^3_x &  \nu u^3_x\\
-\mu u^3_x & 0 &  \phi u^3_x\\
-\nu u^3_x & -\phi u^3_x & 0
\end{pmatrix}
$$
transforms into
$$
\tilde{P}^{ij}=
\begin{pmatrix}
0 & 1 & 0\\
1 & 0 & 0\\
0 & 0 & 0
\end{pmatrix}\frac{d}{dx}+
\begin{pmatrix}
0 & -\tilde\mu u^3_x & \tilde\phi u^3_x\\
\tilde\mu u^3_x & 0 &  \tilde\nu u^3_x\\
-\tilde\phi u^3_x & -\tilde\nu u^3_x & 0
\end{pmatrix}
$$
This means that we can interchange the coefficients $\nu$, $\phi$. Thus, this observation allows us to not consider separately the cases $  \nu=0, \phi\neq 0$ and  $\nu\neq 0, \phi=0$.

If both $\nu$ and $\phi$ vanish, then $\mu=\mu(u^3)$. Suppose $\phi=0$ and $\nu\neq0$ then
$$
\nu=\frac{1}{F-u^2}, \quad \mu=\frac{R}{F-u^2}
$$
where $F$ and $R$ are arbitrary functions depending on $u^3$. Finally, if $\nu\neq0$ and $\phi\neq0$, solving the system we get
$$
\phi=\frac{-Q}{Q u^1+F-u^2}, \quad \nu=\frac{1}{Q u^1+F-u^2},  \quad \mu=\frac{R}{Q u^1+F-u^2}.
$$
These solutions can be summarised as follows:
$$
\begin{array}{c|c|c|c}
\mu =b^{12}_3& \nu=b^{13}_3& \phi =b^{23}_3\\
\hline
R & 0 & 0 & (S.1)\\
\frac{R }{F -u^2} &  \frac{1}{F -u^2}  &0& (S.2)\\
\frac{R}{Q u^1 + F -u^2} &  \frac{1}{Q u^1 + F -u^2} & \frac{-Q}{Q u^1 + F -u^2}  & (S.3)
\end{array}
$$
where $F,R, Q$ are functions depending on $u^3$ and $Q\neq0$.

As shown above in Example \ref{example1}, the restricted admissible transformations are given by
\begin{gather}
u^1= \kappa \tilde u^1+ V(\tilde u^3), \quad u^2= \frac{1}{ \kappa}\tilde u^2+ W(\tilde u^3), \quad u^3=Z(\tilde u^3), \label{change3c_r2}\\
u^1= \kappa \tilde u^2+ V(\tilde u^3), \quad u^3= \frac{1}{\kappa}\tilde u^1+ W(\tilde u^3), \quad u^3=Z(\tilde u^3), \label{change3c_r2_1}
\end{gather}
where $\kappa=const$.

\noindent
{\bf (S.1)}. If $R\neq0$, using the transformation $u^1=\tilde{u}^1$, $u^2=\tilde{u}^2$, $u^3=Z(\tilde{u}^3)$, in the new coordinates $\phi$ transforms into
$\tilde{\phi}=Z_3 F(Z)$. Thus we can always choose $Z$ such that $\phi$ is reduced to 1, obtaining:
\begin{equation}\label{rank2_admiss}
P^{ij}=
\begin{pmatrix}
0 & d_x +u^3_x& 0\\
d_x - u^3_x& 0 & 0\\
0 & 0 & 0
\end{pmatrix}.
\end{equation}
Otherwise, we have the constant coefficient form $\eqref{rank2}_1$,
$$
P^{ij}=
\begin{pmatrix}
0 & d_x & 0\\
d_x & 0 & 0\\
0 & 0 & 0
\end{pmatrix}.
$$
Let us point out that modulo admissible transformations these two structures are equivalent. Indeed, it is sufficient to consider the admissible (but not restricted) change of coordinates
$
\tilde{u}^1=e^{u^3} u^1, \tilde{u}^2=e^{-u^3} u^2, \tilde{u}^3= u^3, 
$
which brings \eqref{rank2_admiss} to $\eqref{rank2}_1$.

\noindent
{\bf (S.2)}. If $R\neq0$, by a transformation of the form \eqref{change3c_r2} with $\kappa=1$, the coefficients $\mu$ and $\nu$ transform as
$$
\tilde \mu=\frac{R(Z) Z_3 -W_3}{F(Z)-W-\tilde{u}^2} , \quad \tilde \nu=\frac{1}{F(Z)-W-\tilde{u}^2}.
$$
Thus, we can choose $W$ and $Z$ such that $\tilde \mu=0$ and $\tilde \nu= -\frac{1}{\tilde{u}^2}$. This leads to the operator of the form $\eqref{rank2}_2$,
$$
P^{ij}=
\begin{pmatrix}
0 & d_x & -\frac{u^3_x}{u^2}\\
d_x & 0 & 0\\
\frac{u^3_x}{u^2} & 0 & 0
\end{pmatrix}.
$$
In the case $R=0$, a shift of $u^2$ implies the same result.

\noindent
{\bf (S.3)}. A transformation of the form \eqref{change3c_r2} with $\kappa=1$ implies
\begin{eqnarray*}
\tilde{\mu} & = & \frac{R(Z) Z_3- V_3 Q(Z)-W_3}{Q(Z) V+ Q(Z) \tilde{u}^1+F(Z) -W-\tilde{u}^2},\\
\tilde{\nu} & = &\frac{1}{Q(Z) V+ Q(Z) \tilde{u}^1+F(Z) -W-\tilde{u}^2},\\
\tilde{\phi} & = &\frac{-Q(Z)}{Q(Z) V+ Q(Z) \tilde{u}^1+F(Z) -W-\tilde{u}^2}.\\
\end{eqnarray*}
Thus, choosing $Z=Q^{-1}$ we can reduce $Q$ to $\tilde{u}^3$. Now, we can choose $V$ and $W$ such that $\tilde\mu=0$, and the denominators of $\nu$ and $\phi$ become $ \tilde{u}^3 \tilde{u}^1-\tilde{u}^2$. We have obtained  $\eqref{rank2}_3$,
$$
P^{ij}=
\begin{pmatrix}
0 & d_x & \frac{1}{u^3u^1-u^2} u^3_x\\
d_x & 0 & \frac{-u^3}{u^3u^1-u^2} u^3_x\\
\frac{-1}{u^3 u^1-u^2} u^3_x & \frac{u^3}{u^3u^1-u^2} u^3_x & 0
\end{pmatrix}.
$$

Once again, one can check that these three structures are not equivalent modulo admissible changes of coordinates.

\noindent
{\bf Case (b)}. It remains to consider the case where
$$
g^{ij}=
\begin{pmatrix}
1 & 0 & 0\\
0 & 1 & 0\\
0 & 0 & 0
\end{pmatrix},
$$
Also in this case all the coefficients $b^{ij}_k$ must be zero except
$$
b^{12}_3=-b^{21}_3=\mu, \quad b^{13}_3=-b^{31}_3=\nu, \quad b^{23}_3=-b^{32}_3=\phi.
$$
which have to satisfy the relations (given by \eqref{G4}),
\begin{equation}\label{rk2_eq2}
\mu_1=\mu \nu, \quad
\nu_1=\nu^2,\quad
\phi_1=\nu \phi, \quad
\mu_2=\mu \phi, \quad
\nu_2=\nu \phi, \quad
\phi_2= \phi^2
\end{equation}
and (given by \eqref{G5}),
$$
\nu_1 \phi=\phi_1 \nu, \quad \nu_2 \phi=\phi_2 \nu, \quad \phi_2 \mu +\nu_1 \mu=\mu_1 \nu+\mu_2 \phi .
$$
As in the previous case, the last three conditions follows from \eqref{rk2_eq2}. Notice that system \eqref{rk2_eq2} is the same given by \eqref{rk2_eq1} if we interchange $\nu$ with $\phi$. Thus, since the changes of coordinate used before, namely $\tilde{u}^1=u^2, \tilde{u}^2=u^1$ and $u^1=\tilde u^1+ V(\tilde u^3), u^2= \tilde u^2 + W(\tilde u^3), u^3=Z(\tilde u^3)$ are restricted admissible transformation also for this metric, the classification in this case follows from the previous one.

We point out that also in this case the two structures
$$
P=\begin{pmatrix}
d_x & 0 & 0\\
0 & d_x & 0\\
0 & 0 & 0
\end{pmatrix},
\quad
P=\begin{pmatrix}
d_x & u^3_x& 0\\
- u^3_x& d_x & 0\\
0 & 0 & 0
\end{pmatrix},
$$
are equivalent up to admissible change of coordinates. Indeed, it is sufficient to consider
the admissible transformation
$$
\tilde{u}^1=\cos(u^3) u^1+\sin(u^3) u^2, \quad \tilde{u}^2=\sin(u^3) u^1-\cos(u^3) u^2, \quad \tilde{u}^3= u^3, 
$$
to bring the second structure to the first one.

\begin{rmk}
All these results are obtained using real change of variables. Allowing complex changes of dependent variables $u^i$, the metrics of rank two \eqref{metrics_rank2} are equivalent. For instance, it is sufficient to choose the transformation:
$$
u^1 = \frac{\tilde{u}^1+\tilde{u}^2}{\sqrt{2}}, \quad u^2 = \frac{i(\tilde{u}^2-\tilde{u}^1)}{\sqrt{2}}, \quad u^3= \tilde{u}^3.
$$
Even though this change of coordinates transforms real coefficients $b^{ij}_k$ to complex, one can easily see that there exist (restricted admissible) complex transformations which reduce the structures $\eqref{rank2}_{2,3}$ to $\eqref{rank2_complex}_{2,3}$.
\end{rmk}

%%%%%%%%
%%%%%%%%
%%%%%%%%

\addcontentsline{toc}{section}{Appendix B. The $\delta$ formalism}
\section*{Appendix B. The $\delta$ formalism}

In this appendix we recall the main aspects of the $\delta$ formalism for Poisson structure of hydrodynamic type. This formalism allows as to ``easily'' compute skew-symmetry, Jacobi identity (through the Schouten bracket), and the Lie derivative for bivectors of the form
\begin{equation}\label{app_biv1}
P^{ij}_k(x,{\bf u}, {\bf u}_x, \ldots, {\bf u}_{k+1})=\sum_{m=0}^{k+1} A^{ij}_m ({\bf u}, {\bf u}_x, \ldots, {\bf u}_{k+1}) \frac{d^{k+1-m}}{dx^{k+1-m}}.
\end{equation}
We are not going to describe the background theory, neither to give all the definitions or theorems. Here we want to give the main tools we have used to reach our result. For a deeper description of the theory, one can refer to \cite{DZ}.

The first step is to translate the form of the bivector \eqref{app_biv1} into the following form
\begin{equation}\label{app_biv}
\tilde{P}^{ij}_k(x-y,{\bf u}, {\bf u}_x, \ldots, {\bf u}_{k+1})=\sum_{m=0}^{k+1} A^{ij}_m ({\bf u}, {\bf u}_x, \ldots, {\bf u}_{k+1}) \delta^{(k+1-m)}(x-y),
\end{equation}
where $\delta^{(s)} (x-y)$ is the $s$-th derivative of the Dirac delta function $\delta (x-y)$ with respect to $x$, and $\delta^0(x-y)=\delta(x-y)$.
For convenience, from now on we call $P^{ij}_{x,y}$ a bivector $\tilde{P}^{ij}_k$. Using this formalism, the skew-symmetry of $P$ is essentially
$$
P^{ij}_{x,y}=-P^{ji}_{y,x}
$$
while, given two bivectors $P,Q$ of the form \eqref{app_biv}, the Schouten bracket is a trivector given by
\begin{gather*}\label{schouten}
 [P,Q]^{ijk}_{x,y,z} =\\
\frac{\partial P^{ij}_{x,y}}{\partial u^{l}_{(s)}(x)} \partial_x^s Q^{lk}_{x,z}
+ \frac{\partial Q^{ij}_{x,y}}{\partial u^{l}_{(s)}(x)} \partial_x^s P^{lk}_{x,z}
+\frac{\partial P^{ij}_{x,y}}{\partial u^{l}_{(s)}(y)} \partial_y^s Q^{lk}_{y,z}
+ \frac{\partial Q^{ij}_{x,y}}{\partial u^{l}_{(s)}(y)} \partial_y^s P^{lk}_{y,z}
\\
+\frac{\partial P^{ki}_{z,x}}{\partial u^{l}_{(s)}(z)} \partial_z^s Q^{lj}_{z,y}
+ \frac{\partial Q^{ki}_{z,x}}{\partial u^{l}_{(s)}(z)} \partial_z^s P^{lj}_{z,y}
+\frac{\partial P^{ki}_{z,x}}{\partial u^{l}_{(s)}(x)} \partial_x^s Q^{lj}_{x,y}
+ \frac{\partial Q^{ki}_{z,x}}{\partial u^{l}_{(s)}(x)} \partial_x^s P^{lj}_{x,y}
\\
+\frac{\partial P^{jk}_{y,z}}{\partial u^{l}_{(s)}(y)} \partial_y^s Q^{li}_{y,x}
+ \frac{\partial Q^{jk}_{y,z}}{\partial u^{l}_{(s)}(y)} \partial_y^s P^{li}_{y,x}
+\frac{\partial P^{jk}_{y,z}}{\partial u^{l}_{(s)}(z)} \partial_z^s Q^{li}_{z,x}
+ \frac{\partial Q^{jk}_{y,z}}{\partial u^{l}_{(s)}(z)} \partial_z^s P^{li}_{z,x}.
\end{gather*}
Finally, the Lie derivative of $P$ along a vector field $\xi$, defined by
$$
\xi=\sum_{i=1}^n \sum_{s\geq 0}\partial^s_x \xi^i({\bf u}(x), {\bf u}_x(x),...) \frac{\partial}{\partial u^i_{(s)}},
$$
is given by
$$
(\mathrm{Lie}_{\xi}P)^{ij} \! =\! \sum_{k,s}\left( \partial^s_x \xi^k({\bf u}(x),...) \frac{\partial A^{ij}}{\partial u^k_{(s)}(x)}- \frac{\partial \xi^i({\bf u}(x),...)}{\partial u^k_{(s)}(x)} \partial ^s_xA^{kj}-\frac{\partial \xi^j({\bf u}(y),...)}{\partial u^k_{(s)}(y)}\partial^s_y A^{ik}\right).
$$
In order to compute these objects, one needs to use some of the properties of the Dirac $\delta$, in particular
\begin{gather}
\delta^{(s)}(y-x)=(-1)^s \delta^{(s)}(x-y),\label{dirac1}\\
f(y) \, \delta^{(s)}(x-y)=\sum_{m=0}^{s} \binom{s}{m} \frac{d^m}{d x^m} f(x) \, \delta^{(s-m)}(x-y). \label{dirac2}
\end{gather}

\subsubsection*{Sketch of the proof of Lemma \ref{lemma_skew}}
For brevity, let us set $\delta^{(m)}(x-y)= \delta^{(m)}_x$, $ \delta^{(m)}(y-x)= \delta^{(m)}_y$.
We want to compute the skew-symmetry condition for a first-order deformation. Now $P_1$ is given by
\begin{eqnarray*}
P^{ij}_{xy}&=&A^{ij}({\bf u}(x)) \delta''_x+ B^{ij}_k({\bf u}(x)) u^k_x \delta'_x +
C^{ij}_k({\bf u}(x)) u^k_{xx} +\tilde{D}^{ij}_{rk}({\bf u}(x)) u^r_x u^k_x,
\end{eqnarray*}
(summation over repeated indices $r$ and $k$ is assumed).
We have to compute $-P^{ji}_{yx}$. Using the properties \eqref{dirac1} and \eqref{dirac2}, we get
\begin{eqnarray*}
-A^{ji}({\bf u}(y)) \delta''_y&=& -A^{ji}({\bf u}(y)) \delta''_x\\
&=&-A^{ji} ({\bf u}(x))  \delta''(x-y) - 2 \partial_x (A^{ji} ({\bf u}(x)) ) \delta'_x
-  \partial_x^2 (A^{ji} ({\bf u}(x)) ) \delta_x\\
&=&-A^{ji} ({\bf u}(x))  \delta''_x - 2\partial_k A^{ji} ({\bf u}(x)) u^k_x \delta'_x\\
&& \qquad -  \left(\partial_{r}\partial_{k} A^{ji} ({\bf u}(x)) u^r_x u^k_x + \partial_k A^{ji} ({\bf u}(x)) u^k_{xx}  \right)\delta_x,\\
- B^{ji}_k({\bf u}(y)) u^k_y \delta'_y&=& B^{ji}_k({\bf u}(y)) u^k_y \delta'_x\\
&=& B^{ji}_k({\bf u}(x)) u^k_x \delta'_x + \partial_x (B^{ji}_k({\bf u}(x)) u^k_x) \delta_x\\
&=& B^{ji}_k({\bf u}(x)) u^k_x \delta'_x
+ \left( \partial_r B^{ji}_k({\bf u}(x))  u^r_x u^k_x + B^{ji}_k({\bf u}(x)) u^k_{xx}\right)\delta_x,\\
- C^{ji}_k({\bf u}(y)) u^k_{yy}\delta_y&=&- C^{ji}_k({\bf u}(x)) u^k_{xx}\delta_x,\\
- \tilde{D}^{ji}_{rk}({\bf u}(y)) u^r_y u^k_y \delta_y&=&-\tilde{D}^{ji}_{rk}({\bf u}(x)) u^r_x u^k_x \delta_x.
\end{eqnarray*}
Computing $P^{ij}_{xy}=-P^{ji}_{yx}$ and comparing the coefficients in the derivative of $\delta$, we get the the first part of Lemma \ref{lemma_skew}.
The second part of the Lemma, involving second-order deformations, can be proved in the same way.

%%%%%%%%%%%%%
%%%%%%%%%%%%%
%%%%%%%%%%%%%

\addcontentsline{toc}{section}{Appendix C. Computation of deformations}
\section*{Appendix C. Computation of deformations}
First of all let us agree about notation: if a function depends only on one variable, we use the symbol $'$ to express the derivative with respect to that variable; otherwise, if a function depends on more than one variable, we use $\partial_i=\partial/ \partial u^i$. Furthermore, to lighten the notation, the functions $\tilde{D}^{ij}_k$, $\tilde{H}^{ij}_{rk}$ and $\tilde{N}^{ij}_{ark}$ will be written without the symbol tilde. Finally, the subscript $f_x$  means the derivative of $f$ with respect to the independent variable $x$.
\begin{rmk}
All the proofs of Theorems \ref{thm_def1}--\ref{thm_def5} are obtained by direct (and cumbersome) computations. For this reason, we are going to discuss in detail only the proof of Theorem \ref{thm_def1}, while we give just a sketch of the proof for the remaining theorems.
\end{rmk}

\subsubsection*{Proof of Theorem \ref{thm_def1}}
We start with deformations of order 1. Imposing the skew-symmetry conditions given by Lemma \ref{lemma_skew}, the number of unknown functions is 12. In particular, apart from $A^{11}=A^{22}=0$, all the coefficients can be written in terms of $A^{21}$, $B^{ij}_k$, $C^{21}_i$, $D^{21}_{ji}$, for $ i,j,k=1,2$ and $i\ge j$.
The Jacobi condition $[P^{(1)}_0,P_1]=0$ implies
$$
B^{11}_1=B^{22}_1=C^{21}_1=D^{21}_{11}=D^{21}_{12}=0,
$$
$$
B^{21}_1=\partial_1 A^{21}, \quad B^{22}_2=r, \quad C^{21}_2=p, \quad D^{21}_{22}=q,
$$
where $p, q, r$ are functions depending on $u^2$. The bivector $P_1$ reads
$$
P_1=
\begin{pmatrix}
0 & \alpha^{12}\\
-\alpha^{12} & 0
\end{pmatrix}
\frac{d^2}{dx^2}
+
\begin{pmatrix}
\beta^{11} & \beta^{12}\\
\beta^{21} & \beta^{22}
\end{pmatrix}
\frac{d}{dx}
+
\begin{pmatrix}
\gamma^{11} & \gamma^{12}\\
\gamma^{21} & \gamma^{22}
\end{pmatrix},
$$
where
\begin{gather*}
\alpha^{12} = - A^{21}, \quad \beta^{11}= B^{11}_2 u^2_x, \quad \beta^{12}=(B^{21}_2- \partial_2 A^{21}) u^2_x-(A^{21})_x,\\
\beta^{21}= \partial_1 A^{21} u^1_x+B^{21}_2 u^2_x, \quad \beta^{22}= r u^2_x, \quad \gamma^{11}= \frac{1}{2} (B^{11}_2 u^2_x)_x,\\
\gamma^{12}= \left( (B^{21}_2 -\partial_2 A^{21}) u^2_x\right)_x- p u^2_{xx}-q (u^2_{x})^2, \quad
\gamma^{21}= p u^2_{xx}+ q (u^2_x)^2, \quad \gamma^{22}= \frac{1}{2} (r u^2_x)_x.
\end{gather*}
Among all these deformations, we have to exclude those that are obtained by infinitesimal Miura transformation.
In order to do this, we need to take an arbitrary vector field $X$ of degree 1, that is
\begin{equation}\label{vect_field1}
X=
\begin{pmatrix}
X^1\\
X^2
\end{pmatrix}
=
\begin{pmatrix}
X^1_1(u^1,u^2) u^1_x + X^1_2(u^1,u^2) u^2_x\\
X^2_1(u^1,u^2) u^1_x + X^2_2(u^1,u^2) u^2_x
\end{pmatrix}.
\end{equation}
The Lie derivative of $P^{(1)}_0$ among $X$ leads to a bivector $Q$ of the form
\begin{equation}\label{Lie1}
Q=\mathrm{Lie}_X P^{(1)}_0=
\begin{pmatrix}
0 & \phi^{12}\\
-\phi^{12} & 0
\end{pmatrix}
\frac{d^2}{dx^2}
+
\begin{pmatrix}
\eta^{11} & \eta^{12}\\
\eta^{21} & 0
\end{pmatrix}
\frac{d}{dx}
+
\begin{pmatrix}
\mu^{11} & \mu^{12}\\
0 & 0
\end{pmatrix},
\end{equation}
where
\begin{gather*}
\phi^{12} = X^2_1, \quad \eta^{11} = 2 (\partial_2 X^1_1-\partial_1 X^1_2) u^2_x, \quad
\eta^{12} = \left(X^2_1 \right)_x + \left( \partial_2 X^2_1-\partial_1 X^2_2\right)u^2_x\\
\eta^{21} = - \partial_1 X^2_1 u^1_x-\partial_1 X^2_2 u^2_x, \quad
\mu^{11} =  \left(( \partial_2 X^1_1-\partial_1 X^1_2)u^2_{x}\right)_x,\\
\mu^{12} = \left(( \partial_2 X^2_1-\partial_1 X^2_2)u^2_{x}\right)_x.
\end{gather*}
Even if the vector field $X$ depends on four functions, in $Q$ we have only three functions, since $X^1_1$ and $X^1_2$ appear always together as $\partial_2 X^1_1-\partial_1 X^1_2$.
At this point, it is not difficult to see that we can eliminate the part of the deformation which involves the functions  $A^{21}, B^{11}_2, B^{21}_2$: it is sufficient to consider the vector field $X$ in the form such that $A^{21}=-X^2_1$, $B^{11}_2=2 (\partial_2 X^1_1-\partial_1 X^1_2)$, $ B^{21}_2=-\partial_1 X^2_2$.
Therefore, the deformations of order 1 leads to $P=P^{(1)}_0+\epsilon \tilde{P}_1 + \mathcal{O}(\epsilon^2)$, where
$$
\tilde{P}_1=
\begin{pmatrix}
0 & -p u^2_{xx} - q(u^2_x)^2 \\
p u^2_{xx} + q(u^2_x)^2 & r u^2_x \frac{d}{dx} + \frac{1}{2} (r u^2_x)_x
\end{pmatrix}.
$$

Now we consider deformations of order 2. In this case, thanks to Lemma \ref{lemma_skew}, the number of unknown functions of $u^1, u^2$ is 30: all the coefficients can be written in terms of $E^{ij}$, $F^{21}_k$, $G^{ij}_{k}$, $H^{ij}_{lk}$, $L^{21}_{k}$, $M^{21}_{sk}$, $N^{21}_{mlk}$, for $ i,j,k,l,m,s=1,2$ and  $j\le i$ and $m\le  l\le k$. The Jacobi condition $2[P^{(1)}_0, P_2]+[\tilde{P}_1,\tilde{P}_1]=0$ implies
$$
G^{22}_1=H^{22}_{11}=H^{22}_{12}=L^{21}_1=M^{21}_{11}=M^{21}_{12}=N^{21}_{111}=N^{21}_{112}=r=0,
$$
$$
G^{21}_2=\partial_1 E^{21}, \quad H^{21}_{11}=\frac{1}{2} \partial_1 F^{21}_1, \quad H^{21}_{12}=\frac{1}{2} \partial_1 F^{21}_2, \quad
H^{11}_{11}=\frac{1}{4} \partial_1^2 E^{11}+\frac{1}{2} \partial_1 G^{11}_1,
$$
$$
H^{11}_{12}=\frac{1}{4} \partial_1 \partial_2 E^{11}+\frac{1}{2} \partial_2 G^{11}_{1}, \quad N^{21}_{122}=-\frac{1}{3}p q, \quad M^{21}_{21}=-p^2,
$$
$$
E^{22}=e, \quad G^{22}_{2}=g, \quad H^{22}_{22}=h, \quad L^{21}_2=l-2p^2 u^1,
$$
$$
M^{21}_{22}=m-2p(p'+3q)u^1, \quad N^{21}_{222}=n-2(p q'+q^2)u^1,
$$
where $e, g, h, l, m, n$ are functions depending on $u^2$. Then, a generic solution for $P_2$ is given by
$$
P_2=
\begin{pmatrix}
\alpha^{11} & \alpha^{12}\\
\alpha^{21} & \alpha^{22}
\end{pmatrix}
\frac{d^3}{dx^3}
+
\begin{pmatrix}
\beta^{11} & \beta^{12}\\
\beta^{21} & \beta^{22}
\end{pmatrix}
\frac{d^2}{dx^2}
+
\begin{pmatrix}
\gamma^{11} & \gamma^{12}\\
\gamma^{21} & \gamma^{22}
\end{pmatrix}
\frac{d}{dx}
+
\begin{pmatrix}
\eta^{11} & \eta^{12}\\
\eta^{21} & \eta^{22}
\end{pmatrix},
$$
where
\begin{gather*}
\alpha^{11}=E^{11},\quad \alpha^{12}=\alpha^{21}=E^{21},\quad \alpha^{22}= e, \quad
\beta^{11} = \frac{3}{2} (E^{11})_x,\\
\beta^{12} = 3(E^{21})_x - ( F^{21}_1 u^1_x + F^{21}_2 u^2_x), \quad
\beta^{21} = F^{21}_1 u^1_x + F^{21}_2 u^2_x, \quad
\beta^{22} = \frac{3}{2} e' u^2_x,
\end{gather*}
and
\begin{eqnarray*}
\gamma^{11} &=& \left(\frac{1}{4} \partial_1^2 E^{11} + \frac{1}{2} \partial_1 G^{11}_1 \right) (u^1_x)^2
+ \left(\frac{1}{2} \partial_1 \partial_2 E^{11} + \partial_2 G^{11}_1 \right) u^1_x u^2_x\\
&&
+G^{11}_1 u^1_{xx} + G^{11}_2 u^2_{xx}+H^{11}_{22} (u^2_x)^2,\\
\gamma^{12} &=&
(3\partial_2 E^{21} - 2 F^{21}_2 + G^{21}_2) u^2_{xx}+(3 \partial_2^2 E^{21} - 2 \partial_2 F^{21}_2 + H^{21}_{22}) (u^2_x)^2
\\&&
+( 4 \partial_1 E^{21} - 2 F^{21}_1) u^1_{xx}+(6 \partial_1 \partial_2 E^{21} -\partial_1 F^{21}_2 - 2 \partial_2 F^{21}_1) u^1_x u^2_x
\\&&
+\left(3 \partial_1^2 E^{21} -\frac{3}{2} \partial_1 F^{21}_1\right)(u^1_x)^2,
\\
\gamma^{21} &=& \partial_1 E^{21} u^1_{xx} +G^{21}_2 u^2_{xx}+\frac{1}{2} \partial_1 F^{21}_1 (u^1_x)^2 + \partial_1 F^{21}_2 u^1_x u^2_x+ H^{21}_{22} (u^2_x)^2,\\
\gamma^{22} &=& h u^2_{xx}+ g (u^2_x)^2,\\
\eta^{11} &=&
\left( \frac{1}{2}\partial_2^2 G^{11}_1 +\frac{1}{2} \partial_1 H^{11}_{22} -\frac{1}{2} \partial_1 \partial_2^2 E^{11} \right) u^1_x (u^2_x)^2
+\left( \frac{1}{2} G^{11}_1-\frac{1}{4} \partial_1 E^{11}\right) u^1_{xxx}
\\
&&
+\left(\frac{1}{2} \partial_1 G^{11}_2+\frac{1}{2} \partial_2 G^{11}_1-\frac{1}{2} \partial_1 \partial_2 E^{11} \right) u^1_x u^2_{xx}
+\left(\frac{1}{2} G^{11}_2-\frac{1}{4} \partial_2 E^{11}\right) u^2_{xxx}
\\
&&
+\left( \frac{1}{2} \partial_2 G^{11}_2 + H^{11}_{22} -\frac{3}{4} \partial_2^2 E^{11} \right) u^2_x u^2_{xx}
+\left( -\frac{1}{8} \partial_1^3 E^{11} +\frac{1}{4} \partial_1^2 G^{11}_1\right) (u^1_x)^3
\\
&&
+\left(\frac{3}{4} \partial_1 \partial_2 G^{11}_1  -\frac{3}{8} \partial_1^2 \partial_2 E^{11} \right) (u^1_x)^2 u^2_x
+\left( \frac{1}{2} \partial_2 H^{11}_{22}-\frac{1}{4} \partial_2^3 E^{11}  \right) (u^2_x)^3
\\&&
+\left( \partial_2 G^{11}_1-\frac{1}{2}\partial_1 \partial_2 E^{11} \right) u^2_x u^1_{xx}
+\left( \partial_1 G^{11}_1-\frac{1}{2}\partial_1^2 E^{11} \right) u^1_x u^1_{xx},
\\
\eta^{12} &=&
(2 \partial_1 E^{21} - F^{21}_1) u^1_{xxx} + (\partial_2 E^{21} - F^{21}_2 + G^{21}_2 + 2 p^2 u^1 -l) u^2_{xxx}\\
&&+(3 \partial_2^2 E^{21} - 3\partial_2 F^{21}_2 + \partial_2 G^{21}_2 + 2 H^{21}_{22} + 2 p (p'+3q)u^1 -m) u^2_x u^2_{xx}\\
&&+\left(\partial_1^3 E^{21}-\frac{1}{2}\partial_1^2 F^{21}_1 \right) (u^1_x)^3 
+\left(3\partial_1^2 \partial_2 E^{21}-\frac{3}{2}\partial_1\partial_2 F^{21}_1 \right) (u^1_x)^2 u^2_x \\
&&+(4 \partial_1^2 E^{21} - 2 \partial_1 F^{21}_1) u^1_x u^1_{xx}
+(4 \partial_1 \partial_2 E^{21} - 2 \partial_2 F^{21}_1) u^2_x u^1_{xx}\\
&&+(3 \partial_1 \partial_2 E^{21} - \partial_1 F^{21}_2 -\partial_2 F^{21}_1 +\partial_1 G^{21}_2 +p^2) u^1_x u^2_{xx}\\
&&+(3 \partial_1 \partial_2^2 E^{21} -\partial_1 \partial_2 F^{21}_2 -\partial_2^2 F^{21}_1 +\partial_1 H^{21}_{22} + pq) u^1_x (u^2_x)^2\\
&&+(\partial_2^3 E^{21} -\partial_2^2 F^{21}_2 +\partial_2 H^{21}_{22} +2(p q' + q^2) u^1 - n) (u^2_x)^3,
\\
\eta^{21} &=& (l-2p^2 u^1 ) u^2_{xxx} -p^2 u^1_x u^2_{xx} +(m-2p (p'+3q)u^1) u^2_x u^2_{xx} - p q u^1_x (u^2_x)^2\\
&&+ (n -2(p q' +q^2)u^1) (u^2_x)^3,\\
\eta^{22} &=& \left( \frac{1}{2} g -\frac{1}{4} e'  \right) u^2_{xxx}+\left(\frac{1}{2} g' + h  -\frac{3}{4} e'' \right) u^2_{xx}  u^2_x + \left(\frac{1}{2} h'  -\frac{1}{4} e''' \right) (u^2_{x})^3.\\
\end{eqnarray*}

In this case, the action of Miura subgroup of infinitesimal transformations is more complicated. Indeed, we have to exclude deformations given by
$$
Q=\mathrm{Lie}_Y \tilde{P}_1 + \mathrm{Lie}_Z P^{(1)}_0, \quad \mbox{such that} \quad \mathrm{Lie}_Y P^{(1)}_0=0
$$
where now $\tilde{P}_1$ is given by
$$
\tilde{P}_1=
\begin{pmatrix}
0 & -p u^2_{xx} - q(u^2_x)^2 \\
p u^2_{xx} +  q(u^2_x)^2 &0
\end{pmatrix},
$$
since $r=0$.
First of all we have to find $Y$ of degree 1 such that $\mathrm{Lie}_Y P^{(1)}_0=0$, that is, we have to bring the coefficients of \eqref{Lie1} to zero. This leads to
$$
Y^1=\partial_1 W(u^1,u^2) u^1_x +\partial_2 W(u^1,u^2)  u^2_x, \quad Y^2=V(u^2) u^2_x,
$$
for arbitrary functions $V, W$. A generic vector field $Z$ of degree 2 is given by
\begin{equation}\label{vect_field2}
Z=
\begin{pmatrix}
Z^1\\
Z^2
\end{pmatrix}
=
\begin{pmatrix}
Z^1_1 u^1_{xx}+ Z^1_2 (u^1_x)^2 + Z^1_3 u^1_x u^2_x + Z^1_4 (u^2_x)^2 +Z^1_5 u^2_{xx}\\
Z^2_1 u^1_{xx}+ Z^2_2 (u^1_x)^2 + Z^2_3 u^1_x u^2_x + Z^2_4 (u^2_x)^2 +Z^2_5 u^2_{xx}
\end{pmatrix},
\end{equation}
where $Z^i_j$ for $i=1,2$, $j=1,\ldots,5$ are arbitrary functions depending on $u^1, u^2$.
By straightforward computation, $Q=\mathrm{Lie}_Y \tilde{P}_1 + \mathrm{Lie}_Z P^{(1)}_0$ is given by
$$
Q=
\begin{pmatrix}
\mu^{11} & \mu^{12}\\
\mu^{21} & 0
\end{pmatrix}
\frac{d^3}{dx^3}
+
\begin{pmatrix}
\nu^{11} & \nu^{12}\\
\nu^{21} & 0
\end{pmatrix}
\frac{d^2}{dx^2}
+
\begin{pmatrix}
\phi^{11} & \phi^{12}\\
\phi^{21} & 0
\end{pmatrix}
\frac{d}{dx}
+
\begin{pmatrix}
\psi^{11} & \psi^{12}\\
\psi^{21} & 0
\end{pmatrix},
$$
where
\begin{gather*}
\mu^{11}=  - 2 Z^1_1, \quad \mu^{12}=\mu^{21}= - Z^2_1,\quad
\nu^{11}=  -3 (Z^1_1)_x,\\
\nu^{12}=  2 Z^2_2 u^1_x + Z^2_3 u^2_x  -3 (Z^2_1)_x,\quad
\nu^{21}= -2 Z^2_2 u^1_x - Z^2_3 u^2_x,
\end{gather*}
and
\begin{eqnarray*}
\phi^{11}&=& (2 \partial_1 Z^1_2 -3 \partial_1^2 Z^1_1 ) (u^2_1)^2
+ (2 \partial_2 Z^1_3-2q \partial_2 W  - 2 \partial_1 Z^1_4 - 3 \partial_2^2 Z^1_1) (u^2_x)^2
\\&&
+(4 \partial_2 Z^1_2-6 \partial_1 \partial_2 Z^1_1) u^1_x u^2_x
+( 2 Z^1_3-2p \partial_2 W  - 2 \partial_1 Z^1_5 -3 \partial_2 Z^1_1) u^2_{xx}
\\&&
+(4 Z^1_2 -5\partial_1 Z^1_1) u^1_{xx},
\\
\phi^{12}&=&
( 4 Z^2_2-4 \partial_1 Z^2_1) u^1_{xx}+(2 Z^2_3-p V + p \partial_1 W -\partial_1 Z^2_5 - 3 \partial_2 Z^2_1) u^2_{xx}\\
&&
+(4\partial_2 Z^2_2 - 6 \partial_1 \partial_2 Z^2_1 +\partial_1 Z^2_3) u^1_x u^2_x + (3\partial_1 Z^2_z -3 \partial_1^2 Z^2_1) (u^1_x)^2\\
&&
+(q \partial_1 W - qV  -\partial_1 Z^2_4 + 2 \partial_2 Z^2_3 - 3 \partial_2^2 Z^2_1) (u^2_x)^2,
\\
\phi^{21}&=&(q \partial_1 W -q V - \partial_1 Z^2_4) (u^2_x)^2
-\partial_1 Z^2_2 (u^1_x)^2 -\partial_1 Z^2_3 u^1_x u^2_x -\partial_1 Z^2_1 u^1_{xx}
\\
&&
+(p \partial_1 W-p V  -\partial_1 Z^2_5)u^2_{xx},\\
\psi^{11}&=&
(3 \partial_2 Z^1_3 - p \partial_2^2 W  - p V'  - 2 q \partial_2 W -p' \partial_2 W -\partial_1 \partial_2 Z^1_5 - 2 \partial_1 Z^1_4- 3 \partial_2^2 Z^1_1) u^2_x u^2_{xx}
\\&&
+(4\partial_1 Z^1_2-4\partial_1^2 Z^1_1) u^1_x u^1_{xx}
+(2 Z^1_2 -2\partial_1 Z^1_1) u^1_{xxx}
+( 4 \partial_2 Z^1_2-4\partial_1 \partial_2 Z^1_1) u^2_x u^1_{xx}
\\&&
+(\partial_2^2 Z^1_3-q \partial_2^2 W - q' \partial_2 W  -\partial_1 \partial_2 Z^1_4 - \partial_2^3 Z^1_1) (u^2_x)^3
+(\partial_1^2 Z^1_2-\partial_1^3 Z^1_1) (u^1_x)^3
\\&&
+( Z^1_3 -p \partial_2 W -\partial_1 Z^1_5-\partial_2 Z^1_1) u^2_{xxx}
+(3 \partial_1 \partial_2 Z^1_2 - 3 \partial_1^2 \partial_2 Z^1_1) (u^1_x)^2 u^2_x
\\&&
+(\partial_1 \partial_2 Z^1_3 + 2 \partial_2^2 Z^1_2-q \partial_1\partial_2 W -\partial_1^2 Z^1_4 - 3 \partial_1 \partial_2^2 Z^1_1) u^1_x (u^2_x)^2
\\&&
+(\partial_1 Z^1_3- p\partial_1 \partial_2 W -\partial_1^2 Z^1_5 - 3 \partial_1 \partial_2 Z^1_1 + 2 \partial_2 Z^1_2) u^1_x u^2_{xx},
\\
\psi^{12}&=&
(p \partial_1^2 W -\partial_1^2 Z^2_5 +\partial_1 Z^2_3 - 3 \partial_1 \partial_2 Z^2_1 + 2 \partial_2 Z^2_2) u^1_x u^2_{xx}
+(\partial_1^2 Z^2_2-\partial_1^3 Z^2_1) (u^1_x)^3
\\&&
+(\partial_2^2 Z^2_3-q' V + q' \partial_1 W - 2 q V' + q\partial_1 \partial_2 W  - p V''  -\partial_1 \partial_2 Z^2_4 -\partial_2^3 Z^2_1 ) (u^2_x)^3\\
&&
+( Z^2_3-p V + p \partial_1 W -\partial_1 Z^2_5 -\partial_2 Z^2_1) u^2_{xxx}
+(4 \partial_1 Z^2_2-4\partial_1^2 Z^2_1) u^1_x u^1_{xx}
\\&&
+( 4 \partial_2 Z^2_2-4 \partial_1 \partial_2 Z^2_1) u^2_x u^1_{xx}
+(3 \partial_1 \partial_2 Z^2_2-3 \partial_1^2 \partial_2 Z^2_1) (u^1_x)^2 u^2_x
\\&&
+(2 Z^2_2 -2\partial_1 Z^2_1) u^1_{xxx}
+(p' \partial_1 W - p'V - 2 q V  - 3p  V'  - 2 \partial_1 Z^2_4
\\&&
+ 2 q \partial_1 W + p\partial_1 \partial_2 W - \partial_1 \partial_2 Z^2_5 + 3 \partial_2 Z^2_3 - 3 \partial_2^2 Z^2_1) u^2_x u^2_{xx}
\\&&
+(q \partial_1^2 W -\partial_1^2 Z^2_4 +\partial_1\partial_2 Z^2_3 + 2 \partial_2^2 Z^2_2-3\partial_1 \partial_2^2 Z^2_1) u^1_x (u^2_x)^2,
\\
\psi^{21}&=& (V' q + V'' p) (u^2_x)^3 + 2 V' p u^2_x u^2_{xx}.\\
\end{eqnarray*}
Comparing the action of Miura subgroup of infinitesimal transformations with the deformations we have obtained, one can see that all the 9 functions depending on two variables in the deformations can be obtained via infinitesimal Miura transformation. In particular one has to consider the following relations among the vector fields and the coefficients of the deformations:
$$
E^{11}=-2 Z^1_1, \quad E^{21}=-Z^2_1, \quad F^{21}_1= - 2 Z^2_2, \quad F^{21}_2 = - Z^2_3,
$$
$$
G^{11}_1= 4 Z^1_2-5\partial_1 Z^1_5, \quad G^{11}_{2}= 2 Z^1_3- 2 p \partial_2 W - 2 \partial_1 Z^1_5-3 \partial_2 Z^1_1,
$$
$$
G^{21}_{2}= p (\partial_1 W -V) -\partial_1 Z^2_5, \quad H^{21}_{22}= q (\partial_1 W -V) -\partial_1 Z^2_4, 
$$
$$
H^{11}_{22}= 2 \partial_2 Z^1_3- 2 q \partial_2 W- 2 \partial_1 Z^1_4-3 \partial_2^2 Z^1_1.
$$
At this point, $P_2$ reads
$$
P_2=
\begin{pmatrix}
0 & 0\\
0 & \alpha^{22}
\end{pmatrix}
\frac{d^3}{dx^3}
+
\begin{pmatrix}
0 & 0\\
0 & \beta^{22}
\end{pmatrix}
\frac{d^2}{dx^2}
+
\begin{pmatrix}
0& 0\\
0& \gamma^{22}
\end{pmatrix}
\frac{d}{dx}
+
\begin{pmatrix}
0 & \eta^{12}\\
-\eta^{12} & \eta^{22}
\end{pmatrix},
$$
where
\begin{gather*}
\alpha^{22} = e,\quad
\beta^{22} = \frac{3}{2} e' u^2_x,\quad
\gamma^{22}= c u^2_{xx} + e (u^2_{x})^2,\\
\eta^{12}= (2p^2 u^1 - l) u^2_{xxx} + \left( 2 u^1 (p q' + q^2) + p V'' + q V' -n \right) (u^2_x)^3 + p^2 u^1_x u^2_{xx}\\
+\left(2 p u^1 (3q +p') + 2 p V' - m \right)u^2_x u^2_{xx} + pq u^1_x (u^2_x)^2,\\
\eta^{22}=\left(\frac{1}{2} g-\frac{1}{4} e' \right) u^2_{xxx}+ \left(\frac{1}{2} g'-\frac{3}{4} e'' + h \right) u^2_x u^2_{xx}+ \left(\frac{1}{2}h' -\frac{1}{4} e'''\right) (u^2_x)^3.
\end{gather*}
The extra freedom due to the function $V$ allows us to set equal to zero one of the two functions $n$ or $g$.
The theorem is proved.

%%%%%%%%%%
%%%%%%%%%%
%%%%%%%%%%
%%%%%%%%%%
\subsubsection*{Proof of Theorem \ref{thm_def2}}
The proof of Theorem \ref{thm_def2} can be obtained in the same way as Theorem \ref{thm_def1}. Due to the lack of space, we are not going to discuss in detail the part of the proof related to second-order deformations, but we give only a sketch of such proof.

We start with deformations of order 1. As in the previous case, apart from $A^{11}=A^{22}=0$, all the coefficients can be written in terms of $A^{21}$, $B^{ij}_k$, $C^{21}_i$, $D^{21}_{ji}$, for $ i,j,k=1,2$ and $i\ge j$, thanks to the skew-symmetry conditions given by Lemma \ref{lemma_skew}.
The Jacobi condition $[P^{(1)}_0,P_1]=0$ implies
$$
B^{11}_1= B^{22}_1= C^{21}_2=0, \quad B^{21}_1=\partial_1 A^{21}, \quad C^{21}_1=-\frac{ A^{21}}{u^1},
\quad D^{21}_{11}=-\frac{\partial_1 A^{21}}{u^1},
$$
$$
D^{21}_{12}=-\frac{\partial_1 B^{21}_2}{u^1}, \quad B^{22}_2=-\frac{2 A^{21}}{u^1} + \frac{r}{(u^1)^3}, \quad D^{21}_{22}=\frac{ B^{11}_2}{2 u^1} + \frac{s}{(u^1)^3},
$$
where $r, s$ are functions depending on $u^2$. The bivector $P_1$ reads
$$
P_1=
\begin{pmatrix}
0 & \alpha^{12}\\
-\alpha^{12} & 0
\end{pmatrix}
\frac{d^2}{dx^2}
+
\begin{pmatrix}
\beta^{11} & \beta^{12}\\
\beta^{21} & \beta^{22}
\end{pmatrix}
\frac{d}{dx}
+
\begin{pmatrix}
\gamma^{11} & \gamma^{12}\\
\gamma^{21} & \gamma^{22}
\end{pmatrix},
$$
where
\begin{gather*}
\alpha^{12} = - A^{21},\quad
\beta^{11} = B^{11}_2 u^2_x,\quad
\beta^{12}= -\partial_1 A^{21} u^1_x+(B^{21}_2-2 \partial_2 A^{21}) u^2_x,\\
\beta^{21}= \partial_1 A^{21} u^1_x+B^{21}_2 u^2_x,\quad
\beta^{22}= \left(\frac{r}{(u^1)^3}  -\frac{2 A^{21}}{u^1}\right)u^2_x,
\end{gather*}
and
\begin{eqnarray*}
\gamma^{11}&=& \frac{1}{2} \left( B^{11}_2 u^2_{xx} + \partial_1 B^{11}_2 u^1_x u^2_x +\partial_2 B^{11}_2 (u^2_x)^2\right),\\
\gamma^{12}&=&
\frac{\partial_1 A^{21}}{u^1} (u^1_x)^2
+\left(\partial_1 B^{21}_2  -\partial_1 \partial_2 A^{21}  +\frac{B^{21}_2}{u^1}\right) u^1_x u^2_x
+\frac{A^{21}}{u^1} u^1_{xx}\\
&&
+(B^{21}_2 -\partial_2  A^{21}) u^2_{xx}
+\left(  \partial_2 B^{21}_2 -\partial_2^2 A^{21} -\frac{B^{11}_2}{2 u^1}- \frac{s}{(u^1)^3}\right) (u^2_x)^2,\\
\gamma^{21}&=&
 \left(\frac{B^{11}_2}{2 u^1}+ \frac{s}{(u^1)^3}\right) (u^2_x)^2
-\frac{\partial_1 A^{21}}{u^1} (u^1_x)^2
-\frac{B^{21}_2}{u^1} u^1_x u^2_x
-\frac{A^{21}}{u^1} u^1_{xx},
\\
\gamma^{22}&=&
\left(\frac{A^{21}}{(u^1)^2} - \frac{\partial_1 A^{21}}{u^1} - \frac{3 r}{2 (u^1)^4}   \right) u^1_x u^2_x
+\left( \frac{ r}{2 (u^1)^3} - \frac{A^{21}}{u^1} \right) u^2_{xx}
\\&&
+ \left(  \frac{\partial_2 r}{2 (u^1)^3} - \frac{\partial_2 A^{21}}{u^1} \right) (u^2_x)^2.
\end{eqnarray*}
The deformations that can be obtained by infinitesimal Miura transformation are given by
$$
Q=\mathrm{Lie}_X P^{(1)}_0=
\begin{pmatrix}
0 & \phi^{12}\\
-\phi^{12} & 0
\end{pmatrix}
\frac{d^2}{dx^2}
+
\begin{pmatrix}
\eta^{11} & \eta^{12}\\
\eta^{21} & \eta^{22}
\end{pmatrix}
\frac{d}{dx}
+
\begin{pmatrix}
\mu^{11} & \mu^{12}\\
\mu^{21} & \mu^{22}
\end{pmatrix},
$$
where
\begin{eqnarray*}
\phi^{12} &=& X^2_1,\\
\eta^{11} &=&2 \left(\partial_2 X^1_1 -\partial_1 X^1_2- \frac{X^1_2}{u^1} \right) u^2_x,\\
\eta^{12} &=& \partial_1 X^2_1 u^1_x+ \left(2 \partial_2 X^2_1-\partial_1 X^2_2  + \frac{X^1_1 - X^2_2}{u^1}\right) u^2_x,\\
\eta^{21} &=& \left(  \frac{X^1_1 - X^2_2}{u^1}-\partial_1 X^2_2  \right)u^2_x - \partial_1 X^2_1 u^1_x, \\
\eta^{22} &=& \frac{2 X^2_1}{u^1} u^2_x,\\
\mu^{11} &=&
\left( \partial_2^2 X^1_1- \partial_1\partial_2 X^1_2 -\frac{\partial_2 X^1_2}{u^1}\right) (u^2_x)^2
+ \left( \partial_2 X^1_1-\partial_1 X^1_2 -\frac{ X^1_2}{u^1}\right) u^2_{xx}
\\
&&
+ \left(\partial_1 \partial_2 X^1_1-\partial_1^2 X^1_2  -\frac{\partial_1 X^1_2}{u^1}+\frac{X^1_2}{(u^1)^2} \right) u^1_x u^2_x,\\
\mu^{12} &=&
\left( \partial_2^2 X^2_1 - \partial_1\partial_2 X^2_2 + \frac{\partial_1 X^1_2 -  \partial_2 X^2_2}{u^1} +\frac{X^1_2}{(u^1)^2}\right) (u^2_x)^2-\frac{\partial_1 X^2_1}{u^1}(u^1_x)^2
\\&&
+ \left( \partial_1 \partial_2 X^2_1 - \partial_1^2 X^2_2 + \frac{\partial_1 X^1_1 - 2 \partial_1 X^2_2}{u^1} \right) u^1_x u^2_x
-\frac{ X^2_1}{u^1}u^1_{xx}
\\&&
+ \left( \partial_2 X^2_1 - \partial_1X^2_2 + \frac{X^1_1- X^2_2}{u^1} \right) u^2_{xx},\\
\mu^{21} &=& \frac{\partial_1 X^2_1}{u^1} (u^1_x)^2- \left( -\frac{\partial_1 X^2_2}{u^1} + \frac{X^1_1-X^2_2}{(u^1)^2} \right) u^1_x u^2_x +\frac{X^2_1}{u^1} u^1_{xx}
\\&&
+\left( \frac{ \partial_2 X^1_1-\partial_1 X^1_2}{u^1}- \frac{X^1_2}{(u^1)^2}\right) (u^2_x)^2, \\
\mu^{22} &=&\left(\frac{\partial_1 X^2_1}{u^1} - \frac{X^2_1}{(u^1)^2} \right) u^1_x u^2_x + \left( \frac{\partial_2 X^2_1}{u^1}\right) (u^2_x)^2 + \frac{X^2_1}{u^1} u^2_{xx}.
\end{eqnarray*}
Also this time, we can eliminate the part of the deformation which involves the functions  $A^{21}, B^{11}_2, B^{21}_2$. Indeed, it is sufficient to consider the vector field $X$ such that
$$
A^{21}=-X^2_1, \quad B^{11}_2=2 \left( \partial_2 X^1_1- \partial_1 X^1_2- \frac{X^1_2}{u^1}\right), \quad B^{12}_2=\frac{X^1_1-X^2_2}{u^1} -\partial_1 X^2_2.
$$
Thus, the deformations of order 1 leads to $P=P^{(1)}_0+\epsilon \tilde{P}_1 + \mathcal{O}(\epsilon^2)$, where
$$
\tilde{P}_1=
\begin{pmatrix}
0 & -\frac{s}{(u^1)^3} (u^2_x)^2  \\
\frac{s}{(u^1)^3} (u^2_x)^2 & \frac{r}{(u^1)^3} u^2_x \frac{d}{dx}+ \frac{1}{2}\left(\frac{r}{(u^1)^3} u^2_x\right)_x
\end{pmatrix}.
$$

In the case of deformations of order 2, thanks to Lemma \ref{lemma_skew}, all the coefficients can be written in terms of $E^{ij}$, $F^{21}_k$, $G^{ij}_{k}$, $H^{ij}_{lk}$, $L^{21}_{k}$, $M^{21}_{sk}$, $N^{21}_{mlk}$, for $ i,j,k,l,m,s=1,2$ and  $j\le i$ and $m\le  l\le k$. The Jacobi condition $2[P^{(1)}_0, P_2]+[\tilde{P}_1,\tilde{P}_1]=0$ implies
$$
E^{22}=\frac{r^2}{2(u^1)^4}, \quad G^{21}_1=\partial_1 E^{21}, \quad G^{22}_1=-\frac{2 r^2}{(u^1)^5}, \quad G^{22}_2=\frac{q}{(u^1)^4}-\frac{2 E^{21}}{u^1},
$$
$$
G^{21}_{2}=3 F^{21}_2-\frac{\partial_1 E^{11}}{2} - 3 \partial_2 E^{21} + (\partial_1 F^{21}_2 + 2 H^{22}_{22}-\partial_1 \partial_2 E^{21})u^1
+\frac{19 s r}{ 6 (u^1)^4} -\frac{E^{11}}{u^1}+\frac{(u^1)^2 \partial_1 H^{22}_{22}}{2},
$$
$$
H^{21}_{11}=\frac{\partial_1 F^{21}_1}{2}, \quad H^{22}_{11}=\frac{15 r^2}{2 (u^1)^6},
\quad
H^{11}_{12}=\frac{\partial_2 G^{11}_1}{2}+\frac{\partial_1 \partial_2 E^{11}}{4}+ \frac{1}{2 u^1} \left(\frac{3\partial_2 E^{11} }{2} -G^{11}_2\right),
$$
$$
H^{11}_{11}=\frac{\partial_1^2 E^{11}}{4}+\frac{ \partial_1 G^{11}_1}{2},
\quad
H^{22}_{12}=\frac{ E^{21}}{ (u^1)^2}+\frac{\partial_1 E^{21}- F^{21}_1}{u^1} - \frac{1}{2(u^1)^5} \left( \frac{9 r r'}{2}-g\right),
$$
\begin{gather*}
H^{21}_{12}=\frac{1}{u^1} \left( \frac{3 \partial_2 E^{21}}{2} -  F^{21}_2 -\frac{\partial_1 E^{11}}{8}+\frac{G^{11}_1}{4}\right)+\frac{\partial_1 \partial_2 E^{21}}{2}-\frac{u^1 \partial_1 H^{22}_{22}}{4}-H^{22}_{22}
\\
+\frac{E^{11}}{2(u^1)^2}-\frac{19 s r}{12 (u^1)^5},\quad
L^{21}_1=-\frac{E^{21}}{u^1}, \quad L^{21}_2=\frac{2 s r }{3 (u^1)^4},
\end{gather*}
$$
M^{21}_{11}=\frac{\partial_1 E^{21}-F^{21}_1}{u^1},
\quad
M^{21}_{12}=\frac{E^{11}}{(u^1)^2}-\frac{2 s r}{ 3 (u^1)^5}+\frac{1}{u^1} \left(\frac{G^{11}_1}{2}-F^{21}_2 -\frac{\partial_1 E^{11}}{4} \right),
$$
\begin{gather*}
M^{21}_{21}=
\frac{1}{u^1} \left(\frac{\partial_1 E^{11}}{2}+3 \partial_2 E^{21}-3 F^{21}_2 \right)
+\partial_1 \partial_2 E^{21}-\partial_1 F^{21}_2 - 2 H^{22}_{22}-\frac{u^1 \partial_1 H^{22}_{22}}{2}
\\
+\frac{E^{11}}{(u^1)^2}-\frac{31 s r}{6 (u^1)^5},
\quad
M^{21}_{22}=\frac{s r' + 13 r s'}{6 (u^1)^4}+\frac{1}{u^1} \left(\frac{G^{11}_2}{2}-\frac{\partial_2 E^{11}}{4} \right)
\end{gather*}
$$
N^{21}_{122}=\frac{\partial_2 E^{11}}{(u^1)^2}-\frac{sr'+13 rs'}{6 (u^1)^5}+\frac{1}{u^1}\left(\frac{\partial_2 G^{11}_1}{2}-\frac{\partial_1\partial_2 E^{11}}{4}-H^{21}_{22} \right),
\quad N^{21}_{111}=-\frac{\partial_1 F^{21}_1}{2 u^1},
$$
$$
N^{21}_{222}=\frac{2 s^2}{(u^1)^5} +\frac{1}{(u^1)^4} \left(n +\frac{1}{2} \int \left(  (u^1)^2 (\partial_2 G^{11}_2 - H^{11}_{22})\right) \, du^1 \right)
+\frac{1}{u^1}\left(\frac{H^{11}_{22}}{2}-\frac{\partial_2^2 E^{11}}{4} \right),
$$
\begin{gather*}
N^{21}_{112}= \frac{\partial_1 H^{22}_{22}}{6} +\frac{1}{(u^1)^2} \left(\frac{2F^{21}_2}{3} +\frac{5 \partial_1 E^{11}}{24}-\frac{G^{11}_1}{12}-\partial_2 E^{21} \right)
-\frac{E^{11}}{(u^1)^3}+\frac{31 s r}{18 (u^1)^6}
\\
+\frac{1}{u^1}\left(\frac{2 H^{22}_{22}}{3}-\frac{\partial_1 \partial_2 E^{21}}{3} -\frac{\partial_1^2 E^{11}}{24} +\frac{\partial_1 G^{11}_1}{12}\right),
\end{gather*}
where $g, n$ are arbitrary functions depending on $u^2$.
We can exclude deformations given by the action of Miura subgroup of infinitesimal transformations of the form $Q=\mathrm{Lie}_Y \tilde{P}_1 + \mathrm{Lie}_Z P^{(1)}_0$, where $Z$ is a generic vector field of degree 2 \eqref{vect_field2} and $Y$ is a vector field of degree 1 \eqref{vect_field1} satisfying the condition $\mathrm{Lie}_Y P^{(1)}_0=0$, namely
$$
Y=
\begin{pmatrix}
\partial_1^2 W  u^1_{x}+ \left(\partial_1 \partial_2 W - \frac{\partial_2 W-V}{u^1} \right)u^2_x\\
\frac{\partial_1 W}{u^1} u^2_x 
\end{pmatrix},
$$
here $W=W(u^1,u^2)$, $V=V(u^2)$ are arbitrary functions.

Comparing the action of Miura subgroup with the deformations we have obtained, the functions depending on two variables in the deformations can be obtained via infinitesimal Miura transformation. In particular one has to consider the following relations among the vector fields and the coefficients of the deformations:
$$
E^{11}=-2 Z^1_1, \quad E^{21}=-Z^2_1,
\quad
F^{21}_2=-Z^2_3 -\frac{Z^2_5+Z^1_1}{u^1}+\frac{r \partial_1 \partial_2 W}{(u^1)^3}+\frac{r (V - \partial_2 W)}{(u^1)^4},
$$
$$
F^{21}_1 = - 2 Z^2_2, \quad
G^{11}_1=4 Z^1_2-5 \partial_1 Z^1_1,
\quad
G^{11}_2=2 (Z^1_3 - \partial_1 Z^1_5) - 3 \partial_2 Z^1_1 -\frac{2 Z^1_5}{u^1},
$$
$$
H^{11}_{22}=2(\partial_2 Z^1_3 -\partial_1 Z^1_4) -3 \partial_2^2 Z^1_1 -\frac{2 s \partial_1 \partial_2 W}{(u^1)^3}
+\frac{2 s (\partial_2 W - V}{(u^1)^4} + \frac{2(\partial_2 Z^1_5 - 2 Z^1_4)}{u^1},
$$
\begin{gather*}
H^{21}_{22}=
\frac{1}{(u^1)^4}\left( \frac{ r' (V-\partial_2 W)}{2} +r (V' -  \partial_2^2 W)- s \partial_1 G\right)
+\frac{Z^1_3 - 2 (Z^2_4 - \partial_2 Z^1_1)}{u^1}\\
\frac{1}{(u^1)^3} \left(r \partial_1 \partial_2^2 W + s \partial_1^2 W +\frac{r' \partial_1 \partial_2 W}{2}\right)
-\partial_1 Z^2_4,
\end{gather*}
$$
H^{22}_{22}=\frac{3 r (\partial_2 G- F)}{(u^1)^5}-\frac{2 r \partial_1 \partial_2 G}{(u^1)^4}+\frac{2 (Z^2_3 - \partial_2 Z^2_1)}{u^1}.
$$
Furthermore, the function $n(u^2)$ can be reduced to zero choosing an infinitesimal Miura transformation given by
$$
Y^1=Y^2=Z^2=0, \quad Z^1= -\frac{n(u^2)}{(u^1)^2}.
$$
By straightforward computation, this leads to \eqref{deg2def2}.

%%%%%%%%%%
%%%%%%%%%%
%%%%%%%%%%
%%%%%%%%%%
\subsubsection*{Proof of Theorem \ref{thm_def3}}
The skew-symmetry conditions given by Lemma \ref{lemma_skew} reduce the number of unknown functions to 48. In particular, apart from $A^{ii}=0$ for $i=1,2,3$, all the coefficients can be written in terms of $A^{ml}$, $B^{ij}_k$, $C^{ml}_k$, $D^{ml}_{ji}$, for $ i,j,k=1,2$ and $i\ge j$ and $m>l$.
The Jacobi condition $[P^{(1)}_0,P_1]=0$ implies
$$
A^{31}=A^{32}=
B^{31}_1=B^{31}_2=
B^{32}_1=B^{32}_2=
B^{33}_1=B^{33}_2=B^{33}_3=0,
$$
$$
C^{31}_1=C^{31}_2=
C^{32}_1=C^{32}_2=
D^{31}_{11}=D^{31}_{12}=
D^{31}_{22}=
D^{32}_{11}=D^{32}_{12}=D^{32}_{22}=0,
$$
$$
B^{11}_1=2 \partial_2 A^{21}-B^{21}_2-\int (\partial_2 B^{22}_2 + \partial_2 B^{21}_1) \, du^1 - 2 F_2,
\quad B^{32}_3=C^{21}_1 -B^{21}_1-\frac{B^{22}_2}{2},
$$
$$
B^{31}_3=\frac{B^{21}_2}{2} +\int \frac{\partial_2 B^{22}_2 + \partial_2 B^{21}_1}{2} \, du^1- C^{21}_2+F_2,
$$
$$
C^{31}_3=\frac{ B^{21}_2}{2} -\int \frac{ \partial_2 B^{22}_2 + \partial_2 B^{21}_1}{2} \, du^1- C^{21}_2-F_2,
\quad C^{32}_3=C^{21}_1 +\frac{B^{22}_2}{2},
$$
$$
D^{21}_{11}=\partial_1 C^{21}_1-\frac{\partial_1 B^{21}_1}{2}-\frac{\partial_1 B^{22}_2+\partial_2 B^{22}_1}{4},
$$
$$
D^{21}_{12}=\frac{\partial_2 C^{21}_1-\partial_2 B^{22}_2 +\partial_1 C^{21}_2}{2}-\frac{\partial_2 B^{21}_1+\partial_1 B^{21}_2 }{4},
$$
$$
D^{21}_{13}=\frac{\partial_3 C^{21}_1+ F_{1}}{2}-\frac{\partial_2 B^{22}_3}{4}-\int \frac{\partial_1 D^{31}_{33}}{2} \, du^2,
$$
$$
D^{21}_{22}=\partial_2 C^{21}_2 +\frac{ \partial_2 B^{21}_2+\partial_1 B^{11}_2}{4}-\int \frac{\partial_2^2 B^{21}_1+\partial_2^2 B^{22}_2}{4} \, du^1 -\frac{\partial_2^2 A^{21}+\partial_2 F_2 }{2},
$$
$$
D^{21}_{23}=\frac{\partial_3 C^{21}_2 +\partial_2 B^{21}_3-D^{31}_{33}-\partial_2 \partial_3 A^{21}}{2}+\frac{\partial_1 B^{11}_3}{4},
$$
$$
D^{31}_{13}=\frac{\partial_2 B^{21}_1-\partial_2 C^{21}_1}{2}+\frac{\partial_2 B^{22}_2}{4},
\quad
D^{32}_{13}=\frac{\partial_1 C^{21}_1-\partial_1 B^{21}_1}{2}-\frac{\partial_1 B^{22}_2}{4},
$$
$$
D^{31}_{23}=\frac{ \partial_2 B^{21}_2}{4} +\int \frac{\partial_2^2 B^{21}_1+\partial_2^2 B^{22}_2}{4} \, du^1 +\frac{\partial_2 F_{2} -\partial_2 C^{21}_2}{2},
$$
$$
D^{32}_{23}=\frac{\partial_1 C^{21}_2}{2} -\frac{ \partial_2 B^{21}_1 +\partial_2 B^{21}_1+\partial_2 B^{22}_2 }{4},
\quad
D^{32}_{33}=-\int \partial_1 D^{31}_{33} \, du^2 + F_1,
$$
where $F_1=F_1(u^1,u^3)$ and $F_2=F_2(u^2,u^3)$.

The deformations that can be eliminated are given by $Q=\mathrm{Lie}_X P^{(3)}_0$, where each component of the vector field $X=(X^1, X^2, X^3)^t$ is given by $X^i=\sum_{m=1}^3 X^i_m(u^1,u^2,u^3) u^m_x$.
By straightforward computation, comparing the action of Miura subgroup with the deformations we have obtained, we can reduce the functions
$B^{11}_3$, $B^{21}_3$, $B^{22}_3$, $C^{21}_1$, $C^{21}_2$, $C^{21}_3$, $D^{21}_{33}$, $D^{31}_{33}$, to zero, choosing the vector field $X$ such that
$$
B^{11}_3=2 X^1_2, \quad B^{21}_3= X^2_2-X^1_1, \quad B^{22}_3=-2 X^2_1, \quad C^{21}_1=-X^3_1, \quad C^{21}_2=-X^3_2,
$$
$$
C^{21}_3=X^2_2-X^3_3, \quad D^{21}_{33}=\partial_1 X^1_3+\partial_2 X^2_3-\partial_3 X^1_1-\partial_3 X^3_3, \quad D^{31}_{33}=\partial_2 X^3_3.
$$
A suitable choice of the vector field allows also to set the function $F_1$ equal to zero. It is sufficient to consider $X$ such that
$$
X^1=X^3_3 u^1_x +2 \left(\int \partial_3 X^3_3 \, du^1 \right) u^3_x,
\quad X^2=X^3_3 u^2_x,
\quad X^3=X^3_3 u^3_x,
$$
where $X^3_3=-\int F_1 \, du^1$.
At this point the deformations depend on $A^{21}$, $B^{11}_2$, $B^{21}_1$, $B^{21}_2$, $B^{22}_1$, $B^{22}_2$, $F_2$. Remarkably, the function $F_2$ never appears alone, but always as $\int \partial_2 B^{22}_2 \, du^1+ 2 F_2$. Let us introduce a new function $r=r(u^1,u^2, u^3)$ such that this object can be replaced by $\partial_2 r(u^1,u^2, u^3)$, namely
$$
 r(u^1,u^2, u^3)=\int B^{22}_2(u^1,u^2, u^3) \, du^1 + 2 \int F_2(u^2, u^3)\, du^2.
$$
Thus, we have $B^{22}_2=\partial_1 r$, and setting $B^{21}_1=\partial_1s$, $A^{21}=a$, $B^{11}_2=b^{11}_2$, $B^{21}_2=b^{21}_2$, $B^{22}_1=b^{22}_1$, we obtain exactly \eqref{3def0}.

%%%%%%
%%%%%%

\subsubsection*{Proof of Theorem \ref{thm_def4}}
Apart from $A^{ii}$ for $i=1,2,3$, all the coefficients can be written in terms of $A^{ml}$, $B^{ij}_k$, $C^{ml}_k$, $D^{ml}_{ji}$, for $ i,j,k=1,2$ and $i\ge j$ and $m>l$.The Jacobi condition $[P^{(1)}_0,P_1]=0$ implies
$$
B^{11}_1=B^{22}_1=B^{32}_1=B^{33}_1=C^{ij}_1=D^{ij}_{11}=D^{ij}_{12}=D^{ij}_{13}=0\quad \mbox{for} \quad i>j,
$$
$$
B^{21}_1=\partial_1 A^{21}, \quad B^{31}_1=\partial_1 A^{31}, \quad A^{32}=a,
$$
$$
B^{ij}_k=b^{ij}_k \quad \mbox{for} \quad i\geq j \geq2 \quad \mbox{and} \quad k=2,3,
$$
$$
C^{ij}_k=c^{ij}_k \quad \mbox{for} \quad i> j \quad \mbox{and} \quad k=2,3,
$$
$$
D^{ij}_{ml}=e^{ij}_{ml} \quad \mbox{for} \quad i>j \quad \mbox{and} \quad 2\leq m\leq l,
$$
where the functions labelled with lowercase letters $a, b^{ij}_k, c^{ij}_k, e^{ij}_{ml}$ depend on $u^2,u^3$. 

The action of Miura subgroup of infinitesimal transformations allows us to eliminate the arbitrary functions depending on three variables in the deformations. Indeed, it is sufficient to consider the vector field $X$ such that the following relations are satisfied
$$
A^{21}= - X^{2}_1, \quad A^{31}=-X^3_1, \quad B^{i1}_k=-\partial_1 X^i_k \quad \mbox{for} \quad i,k=2,3,
$$
$$
B^{11}_k=2(\partial_k X^1_1 -\partial_1 X^1_k) \quad \mbox{for} \quad k=2,3.
$$
This leads to \eqref{3def1}.

\subsubsection*{Proof of Theorem \ref{thm_def5}}
As we have already seen, all the coefficients can be written in terms of $A^{ml}$, $B^{ij}_k$, $C^{ml}_k$, $D^{ml}_{ji}$, for $ i,j,k=1,2$ and $i\ge j$ and $m>l$. The Jacobi condition $[P^{(1)}_0,P_1]=0$ implies
$$
B^{11}_1=B^{22}_2=B^{33}_1=B^{33}_2=C^{21}_2=C^{31}_1=C^{31}_2=C^{32}_1=C^{32}_2=0,
$$
$$
D^{21}_{22}=D^{31}_{11}=D^{31}_{12}=D^{31}_{13}=D^{32}_{11}=D^{32}_{12}=D^{32}_{13}=D^{31}_{22}=D^{31}_{23}=
D^{32}_{22}=D^{32}_{23}=0
$$
$$
B^{11}_2=2 (\partial_1 A^{21}-B^{21}_1), \quad B^{21}_2=\partial_2 A^{21}-\frac{B^{22}_1}{2}, \quad B^{31}_1=\partial_1 A^{31}, \quad B^{31}_2=\partial_1 A^{32},
$$
$$
B^{32}_1=\partial_2 A^{31}, \quad B^{32}_2=\partial_2 A^{32}, \quad B^{33}_3=b, \quad B^{32}_3=\int \partial_2 B^{31}_3 \, du^1 +\mathcal{B}_3, 
$$
$$
B^{21}_3=\mathcal{B}_1+\mathcal{B}_2+\int \left(\partial_3 B^{21}_1+\frac{\partial_2 B^{11}_3}{2} \right) \, du^1 +\int  \frac{\partial_1 B^{22}_3-\partial_3 B^{22}_1}{2}\, du^2,
$$
$$
C^{21}_1=B^{21}_1-\partial_1 A^{21}, \quad
C^{21}_3=\mathcal{B}_2-\partial_3 A^{21} +\int \left( \partial_3 B^{21}_1+\frac{\partial_2  B^{11}_3}{2} \right) \, du^1 + c^{21},
$$
$$
C^{31}_{3}=c^{31}, \quad C^{32}_{3}=c^{32},
\quad
D^{21}_{11}=\partial_1 B^{21}_1-\partial_1^2 A^{21}, \quad D^{21}_{12}=\partial_2 B^{21}_1-\partial_1\partial_2 A^{21},
$$
$$
D^{21}_{23}= \frac{\partial_2\mathcal{B}_2-\partial_2 \partial_3  A^{21}}{2}+\int \left( \frac{\partial_2\partial_3 B^{21}_1}{2}+\frac{\partial_2^2 B^{11}_3}{4}\right) \, du^1,
$$
$$
D^{21}_{13}= \partial_3 B^{21}_1-\partial_1 \partial_3  A^{21}+\frac{\partial_2 B^{11}_3}{4},
\quad D^{31}_{33}=e^{31}, \quad D^{32}_{33}=e^{32},
$$
$$
D^{21}_{33}=\partial_3 \mathcal{B}_2 -\partial_3^2 A^{21} +\int \left( \frac{\partial_2 \partial_3 B^{11}_3}{2}+\partial_3^2 B^{21}_1 \right) \, du^1 + e^{21},
$$
where $\mathcal{B}_1=\mathcal{B}_1(u^1, u^3)$, $\mathcal{B}_2=\mathcal{B}_2(u^2, u^3)$, $\mathcal{B}_3=\mathcal{B}_3(u^2,u^3)$ and $b$, $e^{ij}$, $c^{ij}$, for $i>j$, are arbitrary functions of $u^3$.

Comparing the action of subgroup of infinitesimal transformations with the deformations we have obtained, the arbitrary functions depending on three variables in the deformations can be obtained via infinitesimal Miura transformation. Indeed, it is sufficient to consider the vector field $X$ such that the following relations are satisfied
$$
A^{21}=X^1_2-X^2_1, \quad A^{31}=-X^3_1, \quad A^{32}=-X^3_2,
\quad
B^{11}_3=2 (\partial_3 X^1_1- \partial_1 X^1_3),
$$
$$
B^{21}_1=2\partial_1 X^1_2 -\partial_2 X^1_1 -\partial_1 X^2_1, \quad B^{22}_1=2 (\partial_1 X^2_2 -\partial_2 X^2_1),
$$
$$
B^{22}_3=2 (\partial_3 X^2_2 -\partial_2 X^2_3), \quad B^{31}_3=-\partial_1 X^3_3. 
$$
Once we have eliminated the functions depending on three variables, a deeper analysis of the Miura subgroup of infinitesimal transformations allows also to bring the three functions $\mathcal{B}_i$, $i=1,2,3$, to zero, choosing a vector field $X$ such that
$$
X^1=- u^3_x \int \mathcal{B}_2 \, du^2, \quad X^2=- u^3_x \int \mathcal{B}_1 \, du^1, \quad X^3=- u^3_x \int \mathcal{B}_3 \, du^2.
$$
By straightforward computation, the deformation leads to
$$
P_1=
\begin{pmatrix}
0 & 0 & 0\\
0 & 0 & 0\\
0 & 0 & \beta^{33}
\end{pmatrix}
\frac{d}{dx}
+
\begin{pmatrix}
0 & -\gamma^{21} & -\gamma^{31}\\
\gamma^{21} & 0 & -\gamma^{32}\\
\gamma^{31} & \gamma^{32} & \gamma^{33}
\end{pmatrix}
$$
with
$$
\beta^{33}= b u^3_x,  \quad \gamma^{33} = \frac{1}{2} \left(b u^3_x \right)_x, \quad \gamma^{ij} = e^{ij} (u^3_x)^2 + c^{ij} u^3_{xx} \quad (i>j), \\
$$
where $b, c^{ij}, e^{ij}$, for $i>j$, are arbitrary functions of $u^3$. Finally, choosing a vector field $X$ such that
$$
X^1=-f(u^3) u^2 u^3_x, \quad X^2=f(u^3) u^1 u^3_x, \quad X^3=0,
$$
we have
$$
\mathrm{Lie}_X P_0=
\begin{pmatrix}
0 & -f' (u^3_x)^2 - f u^3_{xx} & 0\\
f' (u^3_x)^2 + f u^3_{xx} & 0 & 0\\
0 & 0 & 0
\end{pmatrix}.
$$
Thus, the freedom in $f$ allows us to eliminate one of the functions $e^{21}$ or $c^{21}$.

%%%%%%%%%%
%%%%%%%%%%

%%%%%%%%%%

\cleardoublepage

\end{document}